\documentclass[a4paper,11pt]{article}
\pdfoutput=1
\usepackage{amsmath,amssymb,amsthm,amsxtra,overpic,bbm,bbold,bm,epsfig,multirow}
\usepackage{color,ulem,subfigure,hyperref,cite}

\allowdisplaybreaks

\begin{document}
\begin{titlepage}
\begin{center}
{\bf\Large
\boldmath{
Twin Modular $S_4$ with $SU(5)$ GUT
}
} 
\\[12mm]
Stephen~F.~King%
\footnote{E-mail: \texttt{king@soton.ac.uk}} 
and
Ye-Ling~Zhou%
\footnote{E-mail: \texttt{ye-ling.zhou@soton.ac.uk}}
\\[-2mm]
\end{center}
\vspace*{0.50cm}
\centerline{ \it
School of Physics and Astronomy, University of Southampton,}
\centerline{\it
Southampton SO17 1BJ, United Kingdom }

\vspace*{1.20cm}

\begin{abstract}
{\noindent
We discuss the $SU(5)$ grand unified extension of flavour models with multiple modular symmetries. The 
proposed model involves two modular $S_4$ groups, one acting in the charged fermion sector, associated with a modulus field value 
$\tau_T$ with residual
$Z_3^T$ symmetry, and one acting in the right-handed neutrino sector, associated with another 
modulus field value $\tau_{SU}$ with residual $Z_2^{SU}$ symmetry. Quark and lepton mass hierarchies are naturally generated with the help of weightons, which are SM singlet fields, where their non-zero modular weights play the role of Froggatt-Nielsen charges.  
The model predicts TM$_1$ lepton mixing, and neutrinoless double beta decay at rates close to the sensitivity of current and future experiments, for both normal and inverted orderings,
with suppressed corrections from charged lepton mixing due to the triangular form of its Yukawa matrix.
}
\end{abstract}
\end{titlepage}

\section{Introduction}

The Standard Model (SM) of particle physics, though highly successful, does not explain why there are three families of quarks and leptons, and does not account for neutrino mass and mixing. The SM also offers no insight into the pattern of quark and lepton (including neutrino) mass and mixing. Moreover, the three gauge groups of the SM are unrelated, with anomaly cancellation and charge quantisation being unexplained, a seemingly accidental consequence of having complete families of quarks and leptons. 

Promising attempts to go beyond the SM in order to address these shortcomings are often based on family symmetry on the one hand, and grand unified theories (GUTs) \cite{Georgi:1974sy} on the other, where the combination of these two symmetries \cite{King:2017guk}
provides a powerful and constrained framework. Neutrino mass and mixing can be included by the addition of gauge singlet states, the right-handed neutrinos, which together with the type I seesaw mechanism \cite{Minkowski:1977sc, Yanagida:1979as, GellMann:1980vs, Glashow:1979nm, Mohapatra:1979ia, Schechter:1980gr, Schechter:1981cv}, provides an elegant understanding of the extreme smallness of neutrino mass compared to charged fermion masses. However understanding the observed approximate tri-bimaximal (TB) nature of neutrino mixing, in which the lepton mixing matrix has a very symmetric structure with $U_{e3}=0$ and equal non-zero elements in the second and third column,
suggests an underlying discrete non-Abelian family symmetry ~\cite{King:2013eh,King:2017guk}. 
For example, the group $S_4$, with generators $S,T,U$~\footnote{We assume the convention $S^2=T^3=U^2=(ST)^3=(SU)^2=(TU)^2=\mathbf{1}$~\cite{King:2013eh}.}, is capable of enforcing 
partial symmetries in the neutrino and charged lepton sector capable of enforcing TB mixing, with $T$ associated with a $Z_3^T$ subgroup preserved in the charged lepton sector, and $S,U$ associated with $Z_2^S\times Z_2^U$ controlling the neutrino sector.

Following the discovery at reactor experiments that $U_{e3}$
takes a non-zero value \cite{Zyla:2020zbs}, exact tri-bimaximal mixing became excluded, while the larger mixing measured initially by atmospheric and solar neutrino experiments, and later by long baseline oscillation experiments, continue to take the tri-bimaximal form. This observation suggests possible relaxed forms of mixing known as trimaximal (TM) mixing, in which the first or second column of the lepton mixing matrix maintains the tri-bimaximal form, while the third column is unconstrained, allowing a non-zero element $U_{e3}$. There is a phenomenological preference for the first case, called TM$_1$ mixing \cite{Varzielas:2012pa, Luhn:2013vna}, in which follows from $S_4$ where $T$ associated with a $Z_3^T$ subgroup preserved in the charged lepton sector, as usual, while the product $SU$ associated with $Z_2^{SU}$ controlling the neutrino sector.

Despite the motivation from neutrino physics for introducing a discrete non-Abelian family symmetry such as $S_4$, this raises the question of the origin of such symmetry, and also the nature of its spontaneous breaking, with particular subgroups preserved in different sectors of the theory, all of which seems to require a large number of flavon fields whose vaccuum alignment requires further driving fields, plus {\it ad hoc} shaping symmetries \cite{Varzielas:2012pa, Luhn:2013vna}. 
The $S_4$, for example, could originate from a continuous non-Abelian symmetry
\cite{deMedeirosVarzielas:2005qg, Koide:2007sr, Banks:2010zn, Luhn:2011ip, Merle:2011vy, Wu:2012ria, Rachlin:2017rvm, King:2018fke}, or have an extra-dimensional origin \cite{Asaka:2001eh, Altarelli:2006kg, Kobayashi:2006wq, Altarelli:2008bg, Adulpravitchai:2009id, Burrows:2009pi, Adulpravitchai:2010na, Burrows:2010wz, deAnda:2018oik, Kobayashi:2018rad, deAnda:2018yfp, Baur:2019kwi}, being an accidental symmetry of orbifolding  \cite{Kobayashi:2008ih,deAnda:2018oik,Olguin-Trejo:2018wpw, Mutter:2018sra}, or as a subgroup of the so-called modular symmetry  \cite{Ferrara:1989bc,Ferrara:1989qb}. Indeed it has been suggested that a finite subgroup of modular symmetry may be well suited to describe neutrino mass and mixing \cite{Altarelli:2005yx, deAdelhartToorop:2011re, Kobayashi:2019mna}, where modular forms play the role of the vacuum alignments of flavon models, but without requiring flavons or driving fields \cite{Feruglio:2017spp}.

In the modular symmetry approach \cite{Feruglio:2017spp}, 
finite discrete flavour symmetry groups such as $S_4$ emerge as the quotient group of modular group $SL(2,Z)$ over the principal congruence subgroups. Quarks and leptons transform nontrivially under the modular $S_4$, for example, and are assigned to have certain  modular weights. Then Yukawa and mass structures emerge as modular forms, holomorphic functions of the complex modulus $\tau$,
with even (or odd) modular weights. 
The modular form of level $N$ and integer weight $k$ can be arranged into some modular multiplets of the homogeneous finite modular group $\Gamma_N\equiv\overline{\Gamma}/\overline{\Gamma}(N)$ if $k$ is an even number~\cite{Feruglio:2017spp}. 
This has been studied for 
$N=2$~\cite{Kobayashi:2018vbk,Kobayashi:2018wkl,Kobayashi:2019rzp,Okada:2019xqk}, $N=3$~\cite{Feruglio:2017spp,Criado:2018thu,Kobayashi:2018vbk,Kobayashi:2018scp,deAnda:2018ecu,Okada:2018yrn,Kobayashi:2018wkl,Novichkov:2018yse,Nomura:2019jxj,Okada:2019uoy,Nomura:2019yft,Ding:2019zxk,Okada:2019mjf,Nomura:2019lnr,Kobayashi:2019xvz,Asaka:2019vev,Gui-JunDing:2019wap,Zhang:2019ngf,Nomura:2019xsb,Wang:2019xbo,Kobayashi:2019gtp,King:2020qaj,Ding:2020yen,Okada:2020rjb,Nomura:2020opk,Okada:2020brs,Yao:2020qyy,Feruglio:2021dte}, $N=4$~\cite{Penedo:2018nmg,Novichkov:2018ovf,deMedeirosVarzielas:2019cyj,Kobayashi:2019mna,King:2019vhv,Criado:2019tzk,Wang:2019ovr,Gui-JunDing:2019wap,Wang:2020dbp}, $N=5$~\cite{Novichkov:2018nkm,Ding:2019xna,Criado:2019tzk} and $N=7$~\cite{Ding:2020msi}.
We shall be interested in $N=4$ case where $\Gamma_N\simeq S_4$.

In general the complex modulus $\tau$ can take any complex value in the upper half complex plane, but it can also be restricted to take particular values, associated with preserved residual symmetries of the finite modular group, which is referred to as the stabiliser in the framework of finite modular symmetry \cite{Novichkov:2018yse,Novichkov:2018ovf,Gui-JunDing:2019wap,deMedeirosVarzielas:2020kji}. However it is clear from the previous discussion that realistic neutrino models require more than one stabiliser to generate interesting models. This implies that more than one modular symmetry must be considered requiring the framework to be extended to multiple modular symmetries \cite{deMedeirosVarzielas:2019cyj}. In a recent paper \cite{King:2019vhv} we considered an example with 
twin modular $S_4$ symmetries, leading to TM$_1$ mixing. The idea was that one modular $S_4$ controls the 
neutrino sector, associated with a modulus field value $\tau_{SU}$, while  
another $S_4$ acts in the charged lepton sector, associated with a modulus field value $\tau_{T}$. 
Apart from the predictions of TM$_1$ mixing, the model led to a novel neutrino mass sum rule
with stringent lower bounds on neutrino masses close to current limits from neutrinoless double beta decay experiments and cosmology.

Such models, while interesting, hitherto do not address the question of quark mass and mixing, nor the possibility of extending the framework of modular symmetry to GUTs. Indeed there has been progress of both questions in the literature, with modular models of both quark and lepton mass and mixing having been proposed \cite{Okada:2019uoy,Lu:2019vgm,King:2020qaj}. In one of these approaches \cite{King:2020qaj} the quark and lepton mass 
hierarchies were explained in the framework of modular symmetry without introducing any additional symmetry such a the Froggatt-Nielsen (FN) $U(1)$ \cite{Froggatt:1978nt} which is commonly used to account for charged fermion mass hierarchies. It was pointed out that the role of the FN flavon may be played by a singlet under both the gauge symmetry and the finite modular symmetry which carries a non-zero modular weight, called the weighton \cite{King:2020qaj}, leading to natural quark and lepton mass hierarchies.
\footnote{Alternatively, fermion mass hierarchies can arise from the slight breaking of certain residual modular symmetries \cite{Okada:2020ukr,Feruglio:2021dte,Novichkov:2021evw}.} 
The question of quark and lepton mass and mixing is necessarily simultaneously addressed in GUTs, and various $SU(5)$ models have been proposed 
\cite{deAnda:2018ecu,Kobayashi:2019rzp,Du:2020ylx,Zhao:2021jxg,Chen:2021zty}
which include finite modular symmetries which act as family symmetries.
Modular symmetry in the context of $SU(5)$ GUTs was first studied in an $(\Gamma_3\simeq A_4)\times SU(5)$ model
in~\cite{deAnda:2018ecu}, then $(\Gamma_2\simeq S_3)\times SU(5)$~\cite{Kobayashi:2019rzp,Du:2020ylx}, and $(\Gamma_4\simeq S_4)\times SU(5)$~\cite{Zhao:2021jxg}. Most recently a comprehensive study of $(\Gamma_3\simeq A_4)\times SU(5)$ has been peformed
\cite{Chen:2021zty}. All these models consider a single modular group.

In this paper, we shall propose an $SU(5)$ GUT with Trimaximal TM$_1$ neutrino mixing arising from twin modular $S_4$ groups.
The resulting model is based on a similar strategy that we developed for our previous lepton model \cite{King:2019vhv}, and indeed the resulting neutrino mass and Yukawa matrices are identical to those considered previously. However, the quark and charged lepton sectors are quite different, with both necessarily having a non-diagonal structure in order to account for quark mixing, unlike the previous model in which the charged lepton Yukawa matrix was diagonal. Also, the hierarchy of the charged lepton masses which was previously obtained by the tuning of Yukawa couplings, is now accounted for by the weighton fields, associated with the twin $S_4$ symmetries, along with the quark mass hierarchies and quark mixing parameters which of course were not considered at all before, since the previous model was purely a model of leptons. Thus we shall construct a flavour model based on $SU(5)$ with two modular symmetries $S_4^F \times S_4^N$, broken to a single $S_4$ by a bi-triplet scalar, leading to the effective theory invariant under one single $S_4$ but involving two modulus fields. The modulus fields gain different VEVs, leading to the breaking of $S_4$ to a residual $Z_3^T$ in the charged fermions sector, and a residual $Z_2^{SU}$ symmetry in the neutrino sector. We analyse the resulting model numerically and show that, in the quark sector, the correct CKM matrix can be achieved, while in the lepton sector, there are small deviations from the TM$_1$ mixing. These corrections are small, but may be tested in the future high-precision neutrino oscillation experiments.

The layout of the remainder of the paper is as follows.
In section~\ref{sec:2}, we briefly review multiple $S$ modular symmetries as the direct origin of fermion masses and mixing. In section~\ref{sec:3}, we propose the $SU(5)$ GUT model based on $S_4^F \times S_4^N$ with two moduli fields. 
We perform a numerical scan in section~\ref{sec:4} to see how good the model match with experimental data. 
Section~\ref{sec:5} concludes the paper. The 
Appendix~\ref{app:1} includes some background information about the modular group and modular forms of level 4, while 
Appendix~\ref{app:2} gives some details about the vacuum alignment required in our model.

\section{Multiple $S_4$ modular symmetries}\label{sec:2}
We review the approach of using multiple finite modular symmetries to explain flavour mixing, with $\Gamma_4 \simeq S_4$ as the representative in the following discussion.

We give a brief introduction of $S_4$ as a finite modular group here. For more details of $\Gamma_4$ and its connection with the infinite modular group, see  Appendix~\ref{app:1}. In the modular $S_4$ symmetry, each element $\gamma$ acting on the complex modulus $\tau$ (${\rm Im}(\tau)>0$) as a linear fractional transformation:
\begin{eqnarray} \label{eq:modular_transformation}
\gamma:~ \tau \to \gamma \tau = \frac{a \tau + b}{c \tau + d}\,,
\end{eqnarray}
where $a, b, c, d$ are integers mod $4$ and satisfy $ad-bc=1$. 
It is convenient to represent each element of the modular $S_4$ group by a $2 \times 2$ matrix, namely,  
\begin{eqnarray}
S_4 = \left\{ \begin{pmatrix} a & b \\ c & d \end{pmatrix} / (\pm \mathbf{1})\,,~~ a, b, c, d \in \mathbb{Z}_4, ~~  ad-bc=1  \right\} \,.
\end{eqnarray}
The finite modular group has two generators, $S_\tau$ and $T_\tau$, which satisfy $S_\tau^2 = (S_\tau T_\tau)^3 = T_\tau^4= \mathbf{1}$.\footnote{Relaxing the last condition $T_\tau^4= \mathbf{1}$ leads to the infinite modular group $\overline{\Gamma}$. Replacing it with $T_\tau^N= \mathbf{1}$ with $N=2,3,5$ leads to finite modular groups $\Gamma_2\simeq S_3$, $\Gamma_3\simeq A_4$ and $\Gamma_5\simeq A_5$, respectively. For $N>5$, additional restrictions are required to make $\Gamma_N$ finite. For example, at $N=7$, one such restriction is $(S_\tau T_\tau)^4 = \mathbf{1}$, leading to the finite $\Gamma_7 \simeq \Sigma(168)$ \cite{Ding:2020msi}.} They acting on the modulus $\tau$ take the following forms
\begin{eqnarray}
S_\tau:~ \tau \to -\frac{1}{\tau} \,, \hspace{1cm}
T_\tau:~ \tau \to \tau + 1\,,
\end{eqnarray}
respectively. They could be represented by $2 \times 2$ matrices as 
\begin{eqnarray}
S_\tau=\begin{pmatrix} 0 & 1 \\ -1 & 0 \end{pmatrix}\,, \hspace{1cm}
T_\tau=\begin{pmatrix} 1 & 1 \\ 0 & 1 \end{pmatrix} \,.
\end{eqnarray}

We consider flavour model construction in the modular invariance approach where the modular symmetry is regarded as the direct origin of flavour mixing. Each field $\phi_i$, which transforms as an irreducible represenation (irrep) $I_i$ of $S_4$ in the flavour space, has a modular weight $2 k_i$. Under the modular transformation, 
\begin{eqnarray}
\gamma:~ \phi_i (\tau) \to \phi_i (\gamma\tau) = (c \tau + d)^{2k_i} \rho_{I_i} (\gamma) \phi_i (\tau) \,. 
\end{eqnarray}
The superpotential in an ${\cal N}=1$ supersymmetry can be generically written in powers of the fields as 
\begin{eqnarray}
w(\phi_i;\tau) = \sum_n  \sum_{\{i_I, \cdots, i_n\}} 
\left(Y_{I_Y} \phi_{i_1} \cdots \phi_{i_n} \right)_{\mathbf{1}} \,, 
\end{eqnarray}
where the subscript $_{\mathbf{1}}$ is the singlet contraction of the coupling and fields. 
Here, the coupling $Y_{I_Y}$ does not have to be a trivial coefficient, but any modular form with the representation  contributed to the singlet contraction and the modular weight $k_Y$ must satisfy $k_Y + k_1+k_2 + \cdots + k_n = 0$. With these requirements, the superpotential is invariant under any tranformation of $S_4$. Besides, $k_Y$ is required to be an even number, as an intrinsic property of modular forms. 
It transforms as 
\begin{eqnarray}
\gamma:~ Y_{I_Y} (\tau) \to Y_{I_Y} (\gamma\tau) = (c \tau + d)^{2k_Y} \rho_{I_Y} (\gamma) Y_{I_Y} (\tau) 
\end{eqnarray}
under the action of $\gamma$. 

Given suitable arrangements for representations and weights for particles in the flavour space, a modular-invariant flavour model can be constructed by including a set of modular forms $\{Y_{I_Y}\}$ which keep the superpotential invariant under the modular transformation. After the modulus field $\tau$ gains a vacuum expectation value (VEV), a specific flavour texture of Yukawa couplings is achieved. 

We extend the discussion to theories including more modular symmetries. Without loss of generality and for convenience of the latter discussion,  we include two, labeled as $S_4^F$ and $S_4^N$, respectively. Let us assume $S_4^F$ and $S_4^N$ to be independent of each other, and the moduli fields are denoted as $\tau_F$ and $\tau_N$, respectively. Any modular transformation $\gamma_F \times \gamma_N$ in $S_4^F \times S_4^N$ takes the form 
\begin{eqnarray} 
&&\gamma_F \times \gamma_N:~ (\tau_F,\tau_N) \to (\gamma_F \tau_F,\gamma_N \tau_N) = 
\left( \frac{a_F \tau_F + b_F}{c_F \tau_F + d_F}, \frac{a_N \tau_N + b_N}{c_N \tau_N + d_N} \right) \,.
\end{eqnarray}
The superpotential $w(\phi_i;\tau_F, \tau_N)$ invariant in $S_4^F \times S_4^N$ is in general expressed as
\begin{eqnarray}
w(\phi_i;\tau_F, \tau_N) = \sum_n  \sum_{\{i_I, \cdots, i_n\}} 
\left(Y_{(I_{Y,F}, I_{Y,N})} \phi_{i_1} \cdots \phi_{i_n} \right)_{(\mathbf{1}, \mathbf{1})} \,, 
\end{eqnarray}
where the field $\phi_i$ and coupling $Y_{(I_{Y,F}, I_{Y,N})}$ transform as 
\begin{eqnarray} \label{eq:field_form_transformation}
&&\phi_i(\tau_F, \tau_N) \to \phi_i(\gamma_F\tau_F, \gamma_N \tau_N) 
\nonumber\\
&&\quad= (c_F\tau_F \!+\! d_F)^{2k_{i,F}}  (c_N\tau_N \!+\! d_N)^{2k_{i,N}} \rho_{I_{i,F}}(\gamma_F) \phi_i(\tau_F, \tau_N) \rho^T_{I_{i,N}}(\gamma_N)\,, \nonumber\\
&&Y_{(I_{Y,F}, I_{Y,N})}(\tau_F, \tau_N) \to Y_{(I_{Y,F}, I_{Y,N})}(\gamma_F \tau_F, \gamma_N \tau_N) \nonumber\\
&&\quad= (c_F\tau_F \!+\! d_F)^{2k_{Y,F}} (c_N \tau_N \!+\! d_N)^{2k_{Y,N}}
 \rho_{I_{Y,F}}(\gamma_F) Y_{(I_{Y,F}, I_{Y,N})}(\tau_F, \tau_N) \rho^T_{I_{Y,N}}(\gamma_N) \,, 
\end{eqnarray}
respectively. Here, we have arranged $\phi_i$ and $Y_{(I_{Y,F}, I_{Y,N})}$ as matrices such that $\gamma_F$ acts on them vertically and $\gamma_N$ acts on them horizontally. The modular weights $2 k_{Y,F}$ and $2 k_{Y,N}$ are even numbers.

In the perspective of model building, as we have shown in \cite{King:2019vhv}, including two modular symmetries  1) allows the modular symmetries to be broken into different residual symmetries in the charged lepton sector and neutrino sector, respectively, and 2) leads to new lepton flavour mixing patten. In the rest of the paper, we extend the multiple modular invariance approach into the quark sector in the framework of grand unification.

\section{An $SU(5)$ model with twin $S_4$ modular symmetries}\label{sec:3}

\begin{table}[h] 
\begin{center}
\begin{tabular}{| l |c c c c c|}
\hline \hline
Fields & $SU(5)$ & $S_4^F$ & $S_4^N$ & $2k_F$ & $2k_N$\\ 
\hline \hline
$T_1$ & $\mathbf{10}$ & $\mathbf{1}$ & $\mathbf{1}$ & $+4$ & $+2$\\
$T_2$ & $\mathbf{10}$ & $\mathbf{1}$ & $\mathbf{1}$ & $+3$ & $+1$\\
$T_3$ & $\mathbf{10}$ & $\mathbf{1}'$ & $\mathbf{1}$ & $0$ & $0$\\
$F$ & $\overline{\mathbf{5}}$ & $\mathbf{3}$ & $\mathbf{1}$ & $0$ & $+2$\\
$N$ & $\mathbf{10}$ & $\mathbf{1}$ & $\mathbf{3}$ & $0$ & $-2$\\
\hline 
$H_{5}$ & $\mathbf{5}$ & $\mathbf{1}$  & $\mathbf{1}$ & $0$ & $0$ \\
$H_{\bar{5}}$ & $\overline{\mathbf{5}}$ & $\mathbf{1}$  & $\mathbf{1}$ & $0$ & $0$ \\
$H_{\bar{45}}$ & $\overline{\mathbf{45}}$ & $\mathbf{1}$  & $\mathbf{1}$ & $0$ & $0$ \\
\hline 
$\Phi$ & $\mathbf{1}$ & $\mathbf{3}$ & $\mathbf{3}$ & $0$ & $0$ \\
$\phi_1$ & $\mathbf{1}$ & $\mathbf{1}$ & $\mathbf{1}$ & $-1$ & $-1$ \\
$\phi_2$ & $\mathbf{1}$ & $\mathbf{1}$ & $\mathbf{1}$ & $-3$ & $-1$ \\
\hline \hline
\end{tabular}
\end{center}
\caption{Transformation properties of fields in the gauge symmetry $SU(5)$ and modular symmetries $S_4^F$ and $S_4^N$. \label{tab:particle_contents}}
\end{table}

\begin{table}[h!] 
\begin{center}
\begin{tabular}{| l | c c c c|}
\hline \hline
Yukawas / masses &$S_4^F$ & $S_4^N$ & $2k_F$ & $2k_N$\\
\hline \hline
$Y_{\bar{5}}^{(2)}(\tau_F), Y_{\bar{45}}^{(2)}(\tau_F)$ & $\mathbf{3}'$ & $\mathbf{1}$ & $+2$ & 0 \\
$Y_{\bar{5}}^{(4)}(\tau_F), Y_{\bar{45}}^{(4)}(\tau_F)$ & $\mathbf{3}$ & $\mathbf{1}$ & $+4$ & 0 \\
$Y_{\bar{5}}^{\prime(4)}(\tau_F), Y_{\bar{45}}^{\prime(4)}(\tau_F)$ & $\mathbf{3}$ & $\mathbf{1}$ & $+4$ & 0 \\
$Y_{\bar{5}}^{(6)}(\tau_F), Y_{\bar{45}}^{(6)}(\tau_F)$ & $\mathbf{3}$ & $\mathbf{1}$ & $+6$ & 0 \\
$Y_{\bar{5}}^{(8)}(\tau_F), Y_{\bar{45}}^{(8)}(\tau_F)$ & $\mathbf{3}$ & $\mathbf{1}$ & $+8$ & 0 \\
$y_{uu}^{(4)}(\tau_F)$ & $\mathbf{3}$ & $\mathbf{1}$ & $+4$ & 0 \\\hline
$M_{\mathbf{1}}(\tau_N)$ & $\mathbf{1}$ & $\mathbf{1}$ & 0 & $+4$ \\
$M_{\mathbf{2}}(\tau_N)$ & $\mathbf{1}$ & $\mathbf{2}$ & 0 & $+4$ \\
$M_{\mathbf{3}}(\tau_N)$ & $\mathbf{1}$ & $\mathbf{3}$ & 0 & $+4$ \\
\hline \hline 
\end{tabular}
\end{center}
\caption{Transformation properties of Yukawa couplings and right-handed neutrino masses in modular symmetries $S_4^F \times S_4^N$. \label{tab:Yukawa_contents}}
\end{table}

Below we construct a flavour model with two $S_4$ modular symmetries. 
The particle contents and their representation and modular weight assignments are given in Table~\ref{tab:particle_contents}. 
In the gauge space, the SM matter fields belong to $\bar{\mathbf{5}}$- and $\mathbf{10}$-plets and the latter are represented as
\begin{eqnarray}
F =
\left(
\begin{array}{c}
d_R^c \\
d_G^c \\
d_B^c \\ 
e \\
-\nu
\end{array}
\right) \,,\quad
T = \frac{1}{\sqrt{2}}
\left(
\begin{array}{ccccc}
0 & u_B^c & -u_G^c & -u_R & -d_R \\
-u_B^c & 0 & u_R^c & -u_G & -d_G \\
u_G^c & -u_R^c & 0 & -u_B & -d_B \\ 
u_R & u_G & u_B & 0 & -e^c \\
d_R & d_G & d_B & e^c & 0
\end{array}
\right) \,.
\end{eqnarray}
The right-handed neutrino field $N$ is introduced to generate neutrino masses. 
$F$ and $N$ are arranged as triplets in $S_4^F$ and $S_4^N$, respectively. $T_1$, $T_2$ and $T_3$ are arranged as singlets of $S_4^F$ and $S_4^N$ but with different modular weights. 
The SM Higgs is embedded into heavy Higgs multiplets $\mathbf{5}$, $\overline{\mathbf{5}}$ and $\overline{\mathbf{45}}$ of $SO(10)$. They are arranged as trivial singlets in the flavour space. We introduce a bi-triplet scalar $\Phi$, which is supposed to break the twin modular $S_4^F$ and $S_4^N$ symmetries into a single $S_4$ symmetry \cite{King:2019vhv}. We further introduce two scalars $\phi_1$ and $\phi_2$. These particles are singlets in both the gauge and flavour space but take non-trivial modular weights, and therefore named as  weightons by authors in \cite{King:2020qaj}. In the present work, they will be responsible for hierarchies of quark masses and charged lepton masses. 

Superpotential terms to generate down quark and charged lepton masses are $w \supset w_{\bar{5}}+ w_{\bar{45}}$ with 
\begin{eqnarray} \label{eq:superpotential_up}
w_{\bar{5}} &=& 
\left[ 
Y^{(6)}_{\bar{5}}(\tau_F) \tilde{\phi}_1 \tilde{\phi}_2^3 + 
Y^{(4)}_{\bar{5}}(\tau_F) \tilde{\phi}_1^2 \tilde{\phi}_2^2 + 
Y^{(8)}_{\bar{5}}(\tau_F) \tilde{\phi}_2^4  \right] T_1 F H_{\bar{5}}  \,, \nonumber\\
&+&Y^{\prime(4)}_{\bar{5}}(\tau_F) \tilde{\phi}_1 \tilde{\phi}_2^2 T_2 F H_{\bar{5}} + 
Y^{(2)}_{\bar{5}}(\tau_F) \tilde{\phi}_1^2 T_3 F H_{\bar{5}}  \,,
\end{eqnarray}
where $\tilde{\phi}_i \equiv \phi_i/\Lambda$ with $\Lambda$ a dimensionful cut-off flavour scale. 
$w_{\bar{45}}$ takes a similar form as $w_{\bar{5}}$ but $Y^{(2k_F)}_{\bar{5}}$ and $H_{\bar{5}}$ are replaced by $Y^{(2k_F)}_{\bar{45}}$ and $H_{\bar{45}}$, respectively. 
Here, $Y^{(2)}_{\cdots}$, $Y^{(4)}_{\cdots}$, $Y^{(6)}_{\cdots}$, $\cdots$ are triplet (either $\mathbf{3}$ or $\mathbf{3}'$) modular forms of modular weights 2, 4, 6, $\cdots$, respectively. All modular forms which are allowed by the modular invariance are listed in Table~\ref{tab:Yukawa_contents}. We fix the VEV of $\tau_F$ at $\langle \tau_F \rangle = \tau_T \equiv \omega$, which preserve a {\it residual $Z_3^T$ symmetry}. Then, we obtain 
\begin{eqnarray}
&&
Y_{\bar{5}}^{(6)} (\tau_T) = y_{e1,\bar{5}}
\begin{pmatrix}
1\\
0\\
0
\end{pmatrix} \,,~
Y_{\bar{5}}^{(4)} (\tau_T) = y_{e2,\bar{5}}
\begin{pmatrix}
0\\
0\\
1
\end{pmatrix} \,,~
Y_{\bar{5}}^{(8)} (\tau_T) = y_{e3,\bar{5}}
\begin{pmatrix}
0\\
1\\
0
\end{pmatrix} \,,\nonumber\\
&&Y_{\bar{5}}^{\prime(4)} (\tau_T) = y_{\mu,\bar{5}}
\begin{pmatrix}
0\\
0\\
1
\end{pmatrix}\,,~
Y_{\bar{5}}^{(2)} (\tau_T)  = y_{\tau,\bar{5}}
\begin{pmatrix}
0\\
1\\
0
\end{pmatrix}\,, 
\end{eqnarray} 
where $y_{e1,\bar{5}}$, $y_{e2,\bar{5}}$, $y_{e3,\bar{5}}$, $y_{\mu,\bar{5}}$ and $y_{\tau,\bar{5}}$ are overall factors which are free parameters in the model. Similarly, we obtain $Y_{\bar{45}}^{(2k_F)} (\tau_T)$ up to overall factors.
These textures lead to Yukawa coupling matrices for down-type quarks and charged leptons in the LR convention as 
\begin{eqnarray} \label{eq:Y_down}
Y_{d} &=& \begin{pmatrix}
y_{dd} \epsilon_1 \epsilon_2^3 & 
y_{ds} \epsilon_1^2 \epsilon_2^2 & 
y_{db} \epsilon_2^4 \\
0 & y_{ss} \epsilon_1 \epsilon_2^2 & 0 \\
0 & 0 & y_{bb} \epsilon_1^2
\end{pmatrix}^* \,, \nonumber\\
Y_{e} &=& \begin{pmatrix}
y_{ee} \epsilon_1 \epsilon_2^3 & 0 & 0 \\
y_{\mu e} \epsilon_1^2 \epsilon_2^2 & y_{\mu \mu} \epsilon_1 \epsilon_2^2 & 0 \\
y_{\tau e} \epsilon_2^4 & 0 & y_{\tau \tau} \epsilon_1^2
\end{pmatrix}^* \,, 
\end{eqnarray}
where $\epsilon_i = \langle \phi_i \rangle / \Lambda$ (for $i=1,2$),  
\begin{eqnarray}
&& y_{dd} = y_{e1,\bar{5}} s + y_{e1,\bar{45}} c, \quad
y_{ds} =y_{e2,\bar{5}} s + y_{e2,\bar{45}} c, \quad
y_{db} = y_{e3,\bar{5}} s + y_{e3,\bar{45}} c, \nonumber\\
&& y_{ss} = y_{\mu,\bar{5}} s + y_{\mu,\bar{45}} c, \quad\;\;
y_{bb} = y_{\tau,\bar{5}} s + y_{\tau,\bar{45}} c,  \nonumber\\ 
&& y_{ee} = y_{e1,\bar{5}} s - 3 y_{e1,\bar{45}} c, \;
y_{\mu e} = y_{e2,\bar{5}} s - 3 y_{e2,\bar{45}} c, \;
y_{\tau e} = y_{e3,\bar{5}} s - 3 y_{e3,\bar{45}} c, \;\nonumber\\
&& y_{\mu \mu} = y_{\mu,\bar{5}} s - 3 y_{\mu,\bar{45}} c, \;\;
y_{\tau \tau} = y_{\tau,\bar{5}} s - 3 y_{\tau,\bar{45}} c 
\end{eqnarray}
with $s = v_{\bar{5}}/(v_{\bar{5}}^2 + v_{\bar{45}}^2)^{1/2}$, $c = v_{\bar{45}}/(v_{\bar{5}}^2 + v_{\bar{45}}^2)^{1/2}$, and $*$ represents the complex conjugation of matrices as we present the result in the left-right notation. 
The above triangular form of the Yukawa matrices in the LR convention ensures that charged lepton corrections 
to PMNS mixing are very suppressed,
while allowing the down quark matrix to contribute dominantly to Cabibbo mixing at order $\epsilon_1$.

Superpotential terms to generate up-type quark masses are given by
\begin{eqnarray} \label{eq:superpotential_up}
w &\supset& (y_{uu} \tilde{\phi}_1^2 \tilde{\phi}_2^2 + y_{uu}^{(4)}(\tau_F) \tilde{\phi}_2^4) T_1 T_1 H_5 + y_{cc} \tilde{\phi}_2^2 T_2 T_2 H_5 + y_{tt} T_3 T_3 H_5 \nonumber\\
&+& y_{uc} \tilde{\phi}_1 \tilde{\phi}_2^2 T_1 T_2 H_5 + y_{ut} \tilde{\phi}_1 \tilde{\phi}_2 T_1 T_3 H_5 + y_{ct} \tilde{\phi}_2 T_2 T_3 H_5 \,, 
\end{eqnarray}
where $y_{uu}$, $y_{cc}$, $y_{tt}$, $y_{uc}$, $y_{ut}$, and $y_{ut}$ are dimensionless coefficients, and $y_{uu}^{(4)}(\tau_F)$ are singlet modular forms of weight 4 of $S_4^F$, seeing Appendix \ref{app:1}. 
The up-type quark Yukawa coupling matrix is given by
\begin{eqnarray}
Y_u = \begin{pmatrix} 
y_{uu} \epsilon_1^2 \epsilon_2^2 & y_{uc} \epsilon_1 \epsilon_2^2 & y_{ut} \epsilon_1 \epsilon_2 \\
y_{uc} \epsilon_1 \epsilon_2^2 & y_{cc} \epsilon_2^2 & y_{ct} \epsilon_2 \\
y_{ut} \epsilon_1 \epsilon_2 & y_{ct} \epsilon_2 & y_{tt} 
\end{pmatrix}^* \,,
\end{eqnarray}
where $y_{uu}^{(4)}(\tau_T) = 0$ was used. 

In the neutrino sector, the superpotential terms to generate neutrino masses are nothing new but a GUT extension of our former work \cite{King:2019vhv}. They are given by
\begin{eqnarray} \label{eq:superpotential_Majorana}
w &\supset& y_{D} \frac{\Phi}{\Lambda}  (F N)_\mathbf{1} H_{5} + M_{\mathbf{1}}(\tau_N) (N N)_{\mathbf{1}} + M_{\mathbf{2}}(\tau_N) (N N)_{\mathbf{2}} + M_{\mathbf{3}}(\tau_N) (N N)_{\mathbf{3}}\,, 
\end{eqnarray}
where the subscript of a bracket, e.g., $(NN)_{\mathbf{3}}$, represents the irreducible representation contraction of the fields in the bracket. 
$\Phi$ is a bi-triplet of $S_4^F \times S_4^N$. Its VEV is not
responsible for special Yukawa textures for leptons, but used to break two modular $S_4$'s to a single modular $S_4$ symmetry \cite{deMedeirosVarzielas:2019cyj},
\begin{eqnarray}
S_4^F \times S_4^N \to S_4\,.
\end{eqnarray}
The VEV of $\Phi$ takes the form
\begin{eqnarray} \label{eq:vev}
\langle \Phi \rangle =  \begin{pmatrix} 1 & 0 & 0 \\ 0 & 0 & 1 \\ 0 & 1 & 0 \end{pmatrix} v_{\Phi} \,.
\end{eqnarray}
The technique of how to achieve this VEV structure was first developed in \cite{King:2019tbt} and has been applied to achieve the breaking of multiple modular symmetries in \cite{deMedeirosVarzielas:2019cyj}. For details of how to derive it in this model, we refer the reader to Appendix~\ref{app:2} of this paper and \cite{deMedeirosVarzielas:2019cyj}. 
After the breaking $S_4^F \times S_4^N \to S_4$, the resulted Yukawa coupling between left-handed and right-handed neutrinos is given by 
\begin{eqnarray} \label{eq:Dirac_mass}
M_{D} &=& y_{D} \begin{pmatrix} 
1 & 0 & 0 \\
0 & 0 & 1 \\
0 & 1 & 0 
\end{pmatrix} \frac{v_\Phi v_{5}}{\Lambda} \,.
\end{eqnarray}
On the other hand, the Majorana mass matrix for right-handed neutrinos is straightforwardly obtained from the singlet, doublet and triplet  modular forms $M_{\mathbf{1}}$, $M_{\mathbf{2}} = (M_{\mathbf{2},1}, M_{\mathbf{2},2})^T$ and $M_{\mathbf{3}} = (M_{\mathbf{3},1}, M_{\mathbf{3},2}, M_{\mathbf{3},3})^T$, i.e., 
\begin{eqnarray}
M_{R} &=& \begin{pmatrix}
M_{\mathbf{1}} & 0 & 0 \\ 
0 & 0 & M_{\mathbf{1}} \\ 
0 & M_{\mathbf{1}} & 0
\end{pmatrix} + 
\begin{pmatrix}
0 & M_{\mathbf{2},1} & M_{\mathbf{2},2} \\ 
M_{\mathbf{2},1} & M_{\mathbf{2},2} & 0 \\ 
M_{\mathbf{2},2} & 0 & M_{\mathbf{2},1}
\end{pmatrix}
+
\begin{pmatrix}
2 M_{\mathbf{3},1} & -M_{\mathbf{3},3} & -M_{\mathbf{3},2} \\ -M_{\mathbf{3},3} & 2M_{\mathbf{3},2} & -M_{\mathbf{3},1} \\ -M_{\mathbf{3},2} & -M_{\mathbf{3},1} & 2M_{\mathbf{3},3}
\end{pmatrix} \,. 
\end{eqnarray}
By assuming the stabiliser in the neutrino sector at $\langle \tau_N \rangle = \tau_{SU} \equiv -\frac{1}{2}+\frac{i}{2}$. This stabiliser is invariant under the action of $SU$. Namely, we are left with a {\it residual $Z_2^{SU}$ symmetry}. Three modular forms $M_{\mathbf{1}}$, $M_{\mathbf{2}}$ and $M_{\mathbf{3}}$ are left with
\begin{eqnarray}
M_{\mathbf{1}}(\tau_{SU}) = a\,,~
M_{\mathbf{2}}(\tau_{SU}) = b
\begin{pmatrix}
1\\
1
\end{pmatrix} \,,~
M_{\mathbf{3}}(\tau_{SU}) = c
\begin{pmatrix}
\sqrt{2}\\
\sqrt{2} - \sqrt{3} \\
\sqrt{2} + \sqrt{3}
\end{pmatrix}\,,
\end{eqnarray}
where $a$, $b$ and $c$ are complex parameters. 
And the Majorana mass matrix for right-handed neutrinos are written in the form of four matrix patterns, i.e.,
\begin{eqnarray} 
M_R = a \begin{pmatrix}
1 & 0 & 0 \\ 0 & 0 & 1 \\ 0 & 1 & 0
\end{pmatrix} + 
b \begin{pmatrix}
0 & 1 & 1 \\ 1 & 1 & 0 \\ 1 & 0 & 1
\end{pmatrix}
+ c\, \sqrt{2}
\begin{pmatrix}
2 & -1 & -1 \\ -1 & 2 & -1 \\ -1 & -1 & 2
\end{pmatrix}
-c\, \sqrt{3}
\begin{pmatrix}
0 & 1 & -1 \\ 
1 & 2 & 0 \\ -1 & 0 & -2
\end{pmatrix} \,.
\label{eq:MR}
\end{eqnarray}
This is explicitly the same mass matrix as that obtained in \cite{King:2019vhv}. 

Without considering the correction from the charged lepton sector, a mass texture as in Eq.~\eqref{eq:MR}, together with a Dirac neutrino  matrix in Eq.~\eqref{eq:Dirac_mass}, leads to the trimaximal TM$_1$ lepton mixing which preserves the first column on the tri-bimaximal
mixing matrix,
\begin{eqnarray}
U_{\rm TM_1}=\left(
\begin{array}{ccc}
 \frac{2}{\sqrt{6}} & - & - \\
 -\frac{1}{\sqrt{6}} &- &- \\
 -\frac{1}{\sqrt{6}} & - & - \\
\end{array}
\right)\,.
\label{TM1}
\end{eqnarray}
Three equivalent relations are predicted from the TM$_1$ mixing, 
\begin{eqnarray}
\tan \theta_{12} = \frac{1}{\sqrt{2}}\sqrt{1-3s^2_{13}}\ \ \ \ {\rm or} \ \ \ \ 
\sin \theta_{12}= \frac{1}{\sqrt{3}}\frac{\sqrt{1-3s^2_{13}}}{c_{13}} \ \ \ \ {\rm or} \ \ \ \ 
\cos \theta_{12}= \sqrt{\frac{2}{3}}\frac{1}{c_{13}} \,.
\label{eq:t12p}
\end{eqnarray}
More precisely, we can write $U_{\rm TM_1}$ explicitly in the form
\begin{eqnarray}
U_{\rm TM_1} = \begin{pmatrix} 
  \frac{2}{\sqrt{6}} & \frac{1}{\sqrt{3}} & 0 \\
 -\frac{1}{\sqrt{6}} & \frac{1}{\sqrt{3}} & \frac{1}{\sqrt{2}} \\
 -\frac{1}{\sqrt{6}} & \frac{1}{\sqrt{3}} & -\frac{1}{\sqrt{2}} \\
\end{pmatrix} \begin{pmatrix} e^{\alpha_3'} & 0 & 0 \\ 0 & \cos\theta_R e^{i\alpha_1} & \sin\theta_R e^{- i\alpha_2} \\ 0 & -\sin\theta_R e^{i\alpha_2} & \cos\theta_R e^{- i\alpha_1} \end{pmatrix} \,,
\end{eqnarray}
which includes three free parameters $\theta_R$, $\alpha_1$, and $\alpha_2$. The $\alpha'_3$ is not an independent parameter but have a complicated correlation with the rest parameter as seen in \cite{King:2019vhv}. The mixing angles and Dirac-type CP-violating phase are determined to be \cite{Luhn:2013vna}
\begin{eqnarray}
\sin\theta_{13} &=& \frac{\sin \theta_R}{\sqrt{3}} \,, \nonumber\\
\tan\theta_{12} &=& \frac{\cos \theta_R}{\sqrt{2}} \,, \nonumber\\
\tan\theta_{23} &=& \left|\frac{ \cos \theta_R + \sqrt{\frac{2}{3}} e^{i (\alpha_1-\alpha_2)} \sin \theta_R}{ \cos \theta_R- \sqrt{\frac{2}{3}} e^{i (\alpha_1-\alpha_2)} \sin \theta_R} \right| \,, \nonumber\\
\delta &=& {\rm arg}\left[ (5 \cos 2 \theta_R +1) \cos ( \alpha_1 - \alpha_2 )-i (\cos 2 \theta_R + 5) \sin ( \alpha_1 - \alpha_2 ) \right]
 \,.
\end{eqnarray}
From these equation, we see that the CP-violating phase $\delta$ and the angle $\theta_{23}$ are correlated via the phase difference $\alpha_1 - \alpha_2$. In the case of $\theta_R \neq 0$ and maximal atmospheric mixing $\theta_{23}=45^\circ$, the phase $\alpha_1 - \alpha_2$ must equal $\pm 90^\circ$, leading to maximal CP violation $\delta = 90^\circ$ or $270^\circ$. We know that the oscillation data shows a small deviation from the maximal atmospheric mixing. Therefore, we expect $\delta$ has a small deviation from its maximal CP-violating value. This will be checked numerically in the next section. 

The mass matrix in Eq.~\eqref{eq:MR} further constrains coefficients for the third and fourth mass matrix patterns on the right hand side with the ratio $-\sqrt{2/3}$. This is a feature different from classical flavour models without modular symmetry, e.g., in \cite{Luhn:2013vna}, where all parameters in front of different patterns are arbitrary. 
We have obtained a neutrino mass sum rule for 
the light neutrino mass eigenvalues
\begin{eqnarray} \label{eq:mass_sum_rule}
\frac{1}{m_1} &=& \Big| \frac{1}{m_2} \left(\sin^2\theta_R e^{-i2\alpha_2} + \sin 2 \theta_R e^{-i (\alpha _1+\alpha _2)} \right) + 
\frac{1}{m_3} \left(\cos^2\theta_R e^{i2\alpha_1} - \sin 2 \theta_R e^{i(\alpha _1+\alpha _2)} \right) \Big|\,. \nonumber\\
\end{eqnarray}
Furthermore, we predicted $m_{ee}$, appearing as the effective mass in neutrino-less double beta decays, correlated with other mass parameters as
\begin{eqnarray}
m_{ee} &=& 
\Big| \frac{2 m_2 m_3}{ 3 \left(
m_2 (\cos^2\theta_R e^{i2\alpha_1} - \sin 2 \theta_R e^{i(\alpha _1+\alpha _2)} )
+ m_3 ( \sin^2\theta_R e^{-i2\alpha_2} + \sin 2 \theta_R e^{-i (\alpha _1+\alpha _2)} ) \right) 
} \Big.  \nonumber\\
&&+ \Big. \frac{1}{3} ( m_2 \cos ^2 \theta_R e^{2 i \alpha_1} + m_3 \sin^2 \theta_R e^{-2 i \alpha_2} ) \Big| \,,
\end{eqnarray}
which  is beyond those reported in \cite{King:2013psa}.

In the present work, the GUT extension induces small corrections from the non-diagonal charged lepton Yukawa coupling $Y_e$. We will check how large is the deviation from the TM$_1$ mixing numerically in the next section.  

\section{Numerical study}\label{sec:4}

We perform a $\chi^2$ analysis and discuss numerical prediction of the model.
 
The charged fermion Yukawa coupling matrices can be diagonalised via $V_f^T (Y_f Y_f^\dag) V_f^T= (Y^{\rm diag}_f)^2$ for $f=u,d,e$ where 
\begin{eqnarray} \label{eq:diagonalisation}
Y^{\rm diag}_u = {\rm diag}\{ \tilde{y}_u, \tilde{y}_c, \tilde{y}_t \}\,, \nonumber\\
Y^{\rm diag}_d = {\rm diag}\{ \tilde{y}_d, \tilde{y}_s, \tilde{y}_b \}\,,\nonumber\\
Y^{\rm diag}_e = {\rm diag}\{ \tilde{y}_e, \tilde{y}_\mu, \tilde{y}_\tau \}\,. 
\end{eqnarray}
For the eigenvalues of the Yukawa couplings, we use the following numerical results, 
\begin{eqnarray}
&&\hspace{-4mm} \tilde{y}_u = (2.92 \pm 1.81) \times 10^{-6}, \quad\;
\tilde{y}_c = (1.43\pm0.100) \times 10^{-3}, \quad
\tilde{y}_t =  0.534 \pm 0.0341, \nonumber\\
&&\hspace{-4mm} \tilde{y}_d = (4.81\pm1.06)\times10^{-6}, \quad\;
\tilde{y}_s = (9.52\pm1.03)\times10^{-5}, \quad\;\,
\tilde{y}_b =(6.95 \pm 0.175) \times 10^{-3}, \nonumber\\
&&\hspace{-4mm} \tilde{y}_e = (1.97 \pm 0.0236) \times 10^{-6}, \;
\tilde{y}_\mu = (4.16 \pm 0.0497) \times 10^{-4}, \;
\tilde{y}_\tau = (7.07 \pm 0.0727) \times 10^{-3}, 
\end{eqnarray}
These values were calculated at the GUT scale from a minimal SUSY breaking
scenario, with $\tan \beta = 5$, as done in \cite{Antusch:2013jca,Bjorkeroth:2015ora,Okada:2019uoy,King:2020qaj}. 
Varying $\tan \beta$ does not change these values significantly except $\tan \beta$ takes a very large value. Three mixing angles and one CP-violating phase in the CKM mixing matrix are applied from the same literature,
\begin{eqnarray}
\theta^q_{12} = 13.027^\circ \pm 0.0814^\circ, \;
\theta^q_{23} = 2.054^\circ \pm 0.384^\circ, \;
\theta^q_{13} = 0.1802^\circ \pm 0.0281^\circ, \;
\delta^q =69.21^\circ \pm 6.19^\circ.
\end{eqnarray}
For neutrino masses and lepton mixing, we take global best-fit (bf) values (without including SK atmospheric data) from NuFIT 5.0 \cite{Esteban:2020cvm, nufit5} and average the positive and negative $1\sigma$ errors. 
\begin{eqnarray}
&\Delta m^2_{21} = (7.42 \pm 0.21) \times 10^{-5} {\rm eV}^2\,, \;
\Delta m^2_{3l} = (2.514 \pm 0.028) \times 10^{-3} {\rm eV}^2 \,, \nonumber\\
&\theta_{12} = 33.44^\circ \pm 0.77^\circ\,, \qquad
\theta_{23} = 49.0^\circ \pm 1.3^\circ\,, \qquad
\theta_{13} = 8.57^\circ \pm 0.13^\circ\,,
\end{eqnarray}
for the normal ordering (NO, i.e., $m_1 < m_2 < m_3$) of neutrino masses and
\begin{eqnarray}
&\Delta m^2_{21} = (7.42 \pm 0.21) \times 10^{-5} {\rm eV}^2\,, \;
\Delta m^2_{3l} = -(2.497 \pm 0.028) \times 10^{-3} {\rm eV}^2 \,, \nonumber\\
&\theta_{12} = 33.45^\circ \pm 0.77^\circ\,, \qquad
\theta_{23} = 49.3^\circ \pm 1.1^\circ\,, \qquad
\theta_{13} = 8.61^\circ \pm 0.12^\circ\,,
\end{eqnarray}
for the inverted ordering (IO, i.e., $m_2 < m_3 < m_1$), where $\Delta m^2_{3l} = \Delta m^2_{31}$ (for NO) and $\Delta m^2_{32}$ (for IO). 

\begin{figure}[t!]
\centering
\includegraphics[width=.6\textwidth]{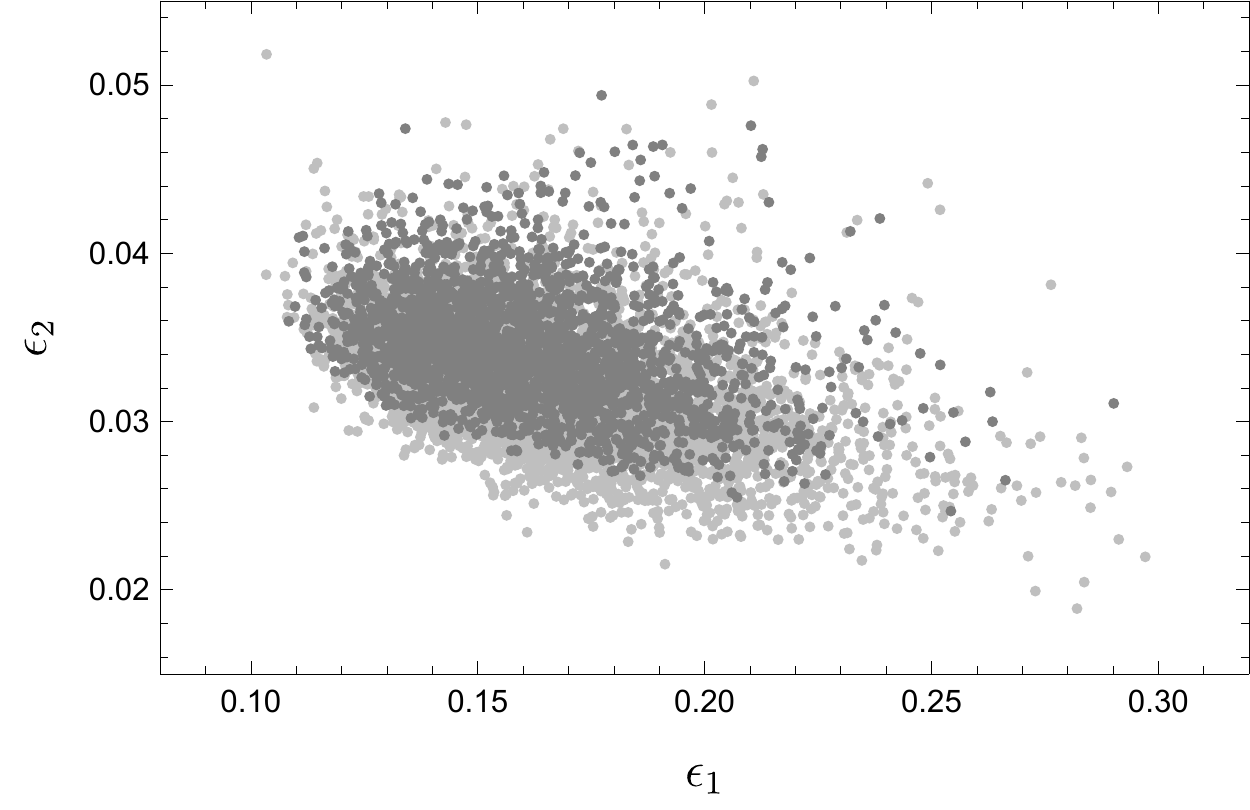}
\caption{A scan plot of $\epsilon_1$ and $\epsilon_2$ with $\chi^2<1$ (gray) and $\chi^2<10$ (light gray), respectively.} \label{fig:epsilons}
\end{figure}

To check the phenomenological viability of the model, we perform a simple $\chi^2$ analysis. In the quark sector, the $\chi^2$ function is defined via
\begin{eqnarray} \label{eq:chi_1}
\chi^2 &=& \sum_{q \in {\rm Obs}} \Big(\frac{q({\rm Para}) - q^{\rm bf}}{\sigma_q}\Big)^2 \,, \nonumber\\
{\rm Para} &=& \{ y_{uu}, y_{uc}, y_{ut}, y_{cc}, y_{ct}, y_{tt}, y_{dd}, y_{ds}, y_{db}, y_{ss}, y_{bb}, \epsilon_1, \epsilon_2 \} \,,\nonumber\\
{\rm Obs} &=& \{ \tilde{y}_u, \tilde{y}_c, \tilde{y}_t, \tilde{y}_d, \tilde{y}_s, \tilde{y}_b, \theta_{12}^q, \theta_{23}^q,  \theta_{13}^q, \theta^q\} \,. 
\end{eqnarray}
Here, in the Para set, $\epsilon_1$ and $\epsilon_2$  are scanned in the range $(0,0.3)$ and all coefficients $y_{\alpha\beta}$ are scanned with absolute values in the range $(0,1)$ and phases in $(0, 2\pi)$. Due to the large parameter space, the model can fit the experimental data very well. 
We have checked that in all points with small $\chi^2$, a hierarchy with $\epsilon_1 \sim \sqrt{\epsilon_2} \sim {\cal}O(\theta_C/\sqrt{2})$ is obtained, where $\theta_C \approx 0.2276$ is the Cabibbo angle. A scan plot for $\epsilon_1$ and $\epsilon_2$ with $\chi^2<1$ (blue) and $\chi^2<10$ (light blue), respectively, is shown in Fig.~\ref{fig:epsilons}. 
Below is a sample of parameters and predictions of observables with $\chi^2 = 0.001$, 
\begin{eqnarray} \label{eq:para_quark}
&y_{uu} =0.0681+0.3260i,~
y_{uc} =0.4531-0.2645i,~
y_{ut} =-0.2657-0.2039i, \nonumber\\
&y_{cc} =-0.3886-1.0082i,~
y_{ct} =-0.5947+0.0148i,~
y_{tt} =-0.4844-0.2270i,  \nonumber\\
&y_{dd} =0.7593+0.5024i,~
y_{ds} =0.6024-0.1092i,~
y_{db} =0.6070+0.1213i, \nonumber\\
&y_{ss} =0.5315+0.1975i,~
y_{bb} =0.2788+0.8131i, \nonumber\\
&\epsilon_1 = 0.1553,~
\epsilon_2 =0.03259
\end{eqnarray}
and 
\begin{eqnarray}
&\tilde{y}_d = 4.829 \times 10^{-6},~
\tilde{y}_s = 9.486 \times 10^{-5},~
\tilde{y}_b = 6.943 \times 10^{-3},\nonumber\\
&\tilde{y}_u = 2.901 \times 10^{-6},~
\tilde{y}_c = 1.428 \times 10^{-3},~
\tilde{y}_t = 0.5357, \nonumber\\
&\theta_{12}^q = 13.02^\circ,~
\theta_{23}^q = 2.071^\circ,~
\theta_{13}^q = 0.1807^\circ,~
\delta^q = 69.33^\circ .
\end{eqnarray}

\begin{figure}[t!]
\centering
\includegraphics[width=.47\textwidth]{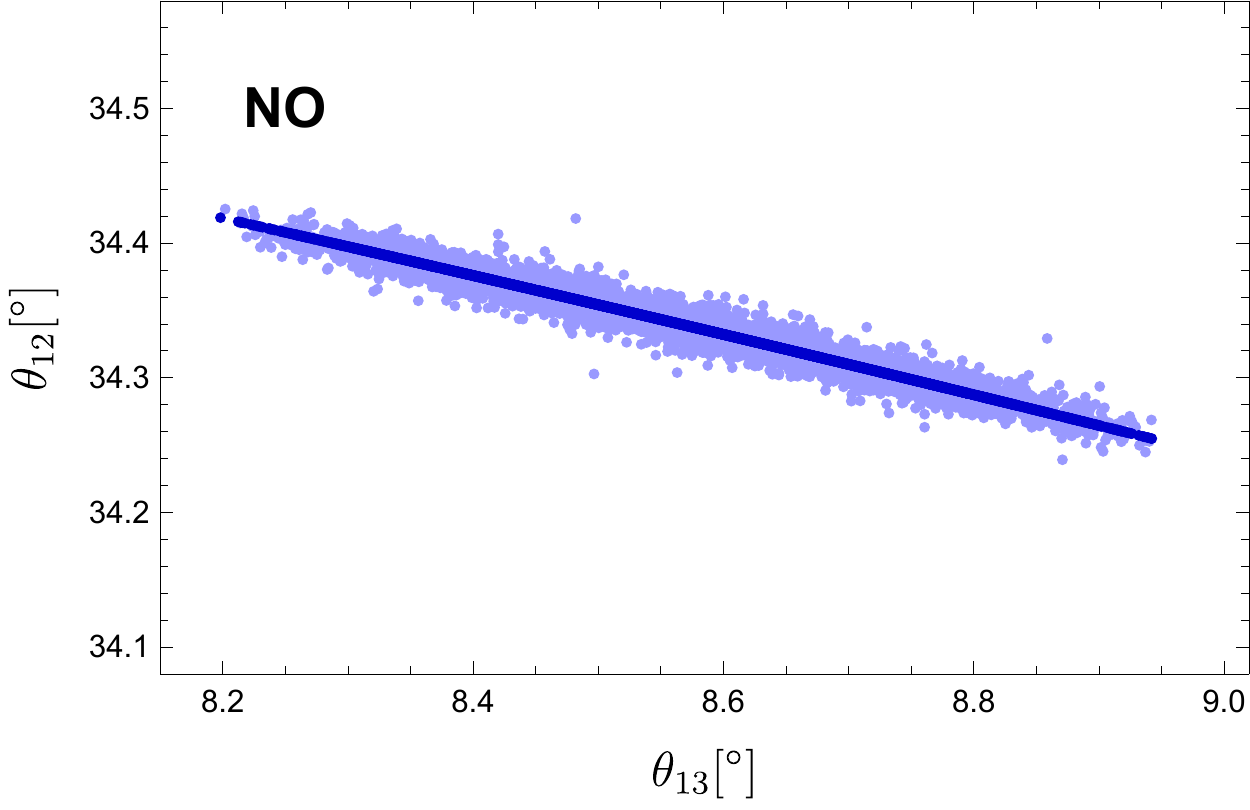}
\includegraphics[width=.47\textwidth]{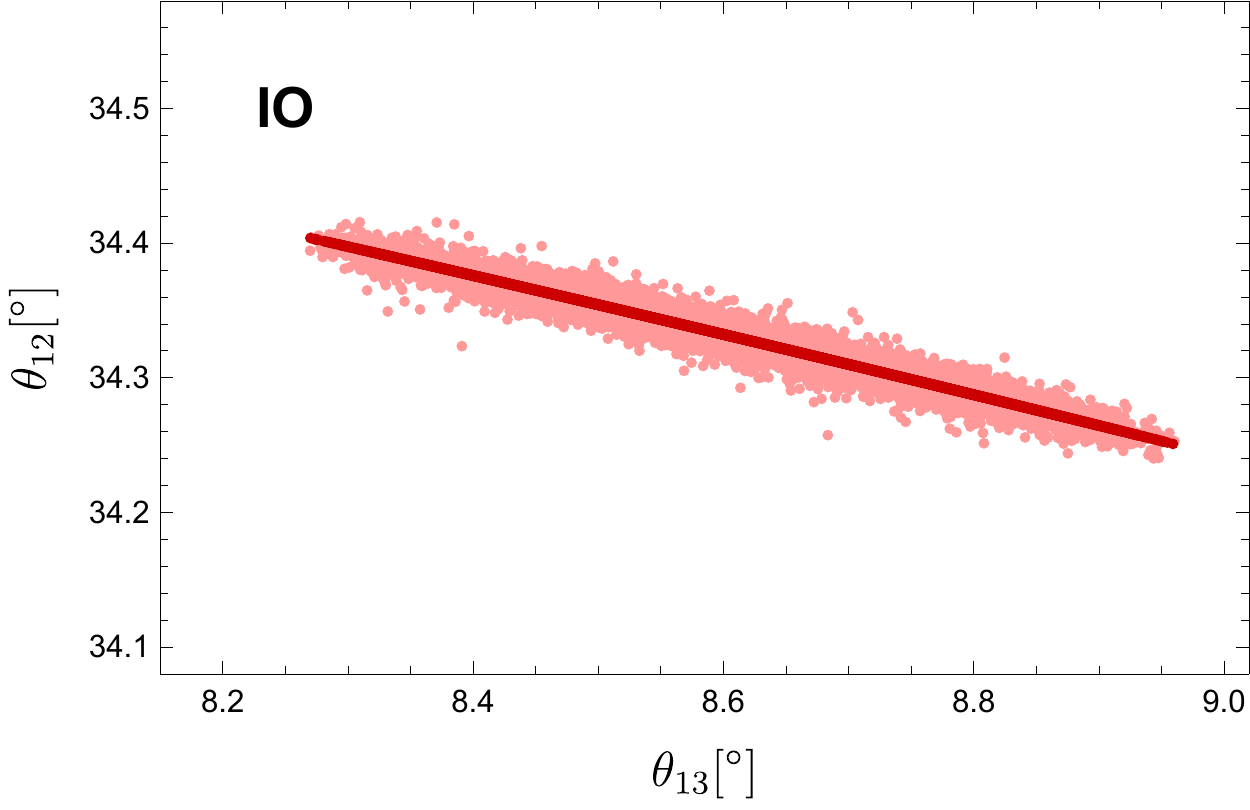}
\caption{Prediction of lepton mixing angles with $\chi^2<10$ in the normal ordering (NO, left panel) and inverted ordering (IO, right panel) of neutrino masses. Results with and without the correction from the charged lepton sector are shown. The latter predicts the TM$_1$ mixing explicitly, as shown in the dark red/blue curves. 
} \label{fig:mixing}
\end{figure}
\begin{figure}[h!]
\centering
\includegraphics[width=.47\textwidth]{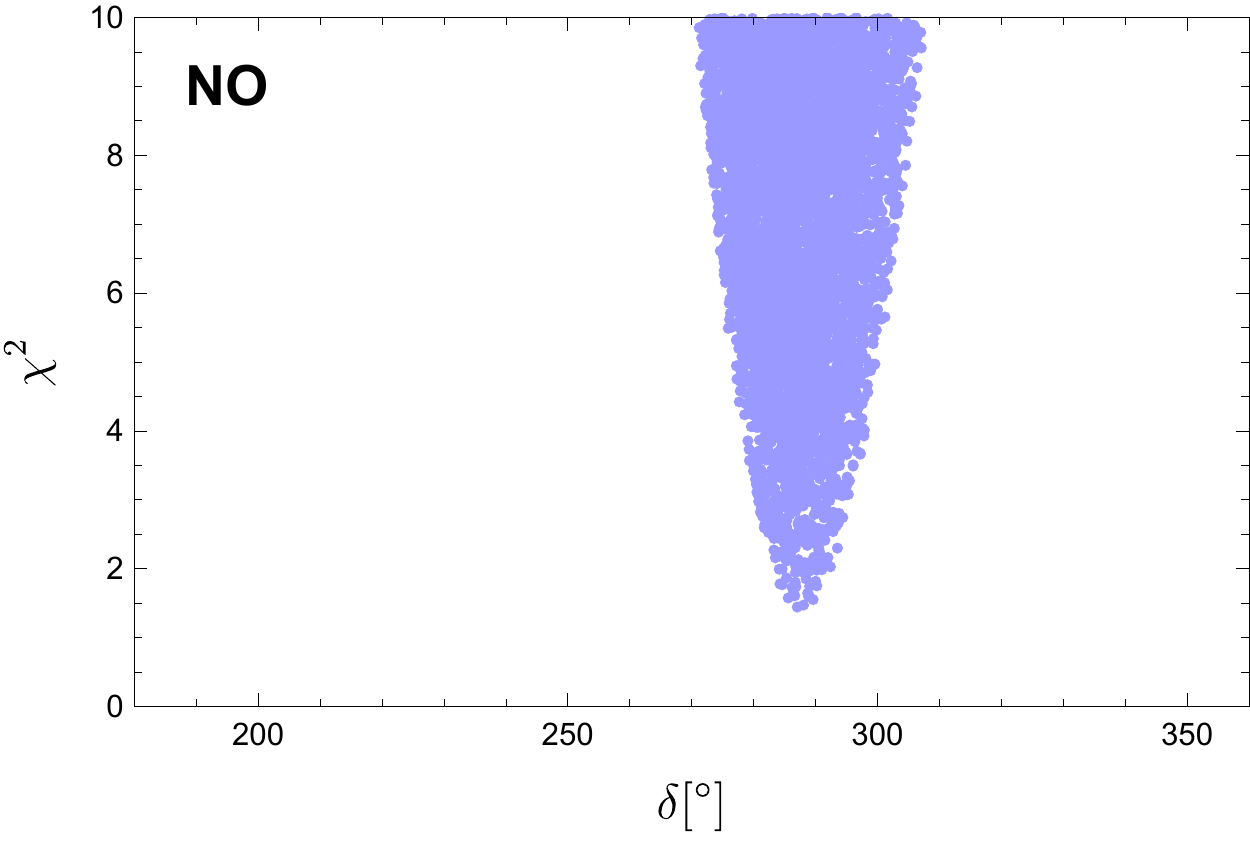}
\includegraphics[width=.47\textwidth]{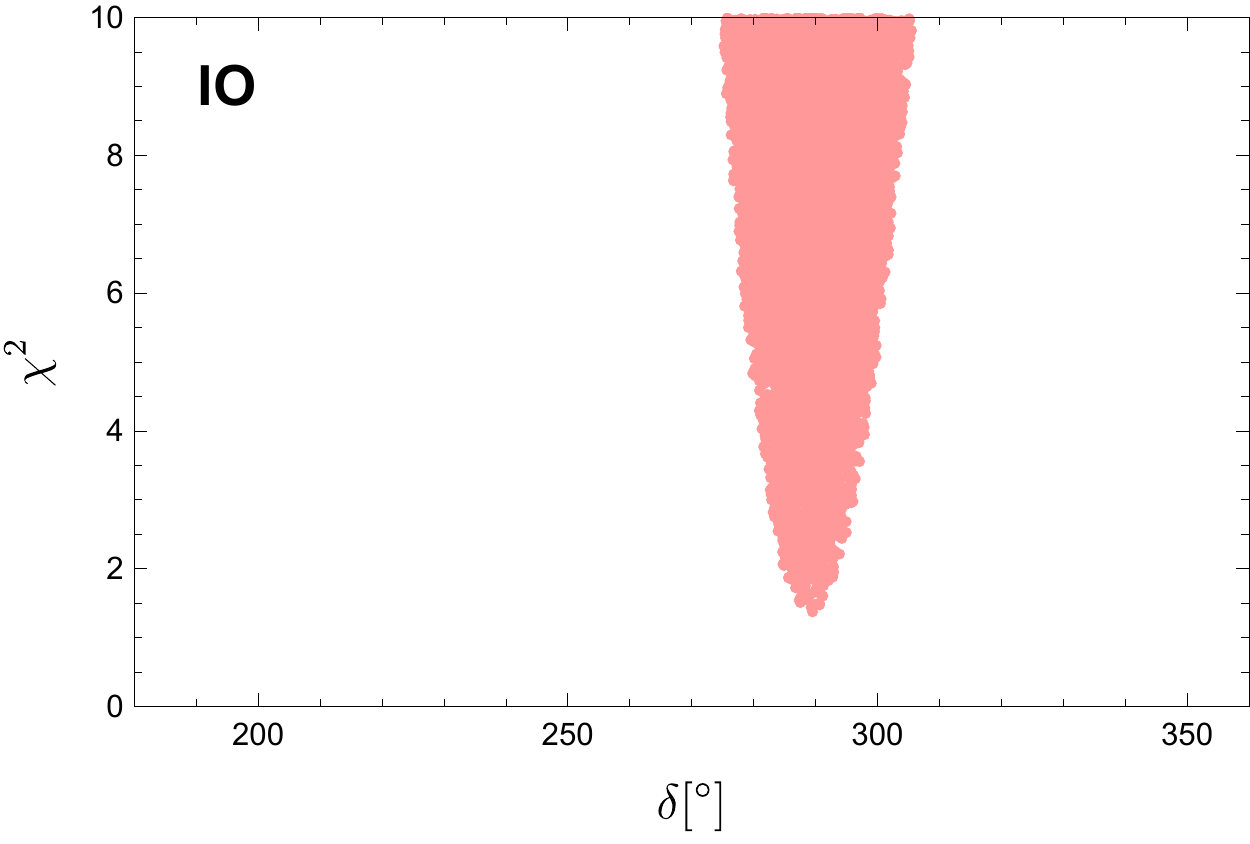}
\caption{Prediction of the CP-violating phase $\delta$ with $\chi^2<10$. Same inputs as in Fig.~\ref{fig:mixing} are used.} \label{fig:delta_chisq}
\end{figure}

We discuss the phenomenological prediction in the lepton sector. We have recovered the same neutrino mass matrix as in \cite{King:2019vhv}. Such a texture leads to the TM$_1$ mixing if the charged lepton mass matrix is diagonal. However, as we see in Eq.~\eqref{eq:Y_down}, some small off-diagonal entries of $Y_l$ are allowed by the modular invariance. Therefore, deviations from the TM$_1$ mixing are expected. Off-diagonal entries of the unitary matrix $V_e$, which is used to diagonalise $Y_e Y_e^\dag$ (c.f. Eq.~\eqref{eq:diagonalisation}), are estimated to be 
$(V_e)_{12} \sim \epsilon_1 \epsilon_2$, $(V_e)_{23} \sim \epsilon_2^6/\epsilon_1^2$ and $(V_e)_{13} \sim \epsilon_2^7/\epsilon_1^3$. By taking values of $\epsilon_1$ and $\epsilon_2$ in Fig.~\ref{fig:epsilons}, these entries can maximally reach  
$(V_e)_{12} \sim 4 \times 10^{-3}$, and $(V_e)_{23,13} \sim 10^{-5}$. Therefore, a deviation less than one percent may be induced by $(V_e)_{12}$. We perform a $\chi^2$ analysis to compare the prediction with and without the correction from charged lepton sector. 
The $\chi^2$ function is defined following Eq.~\eqref{eq:chi_1}, but the set of Obs and Para are given differently, 
\begin{eqnarray} \label{eq:chi_2}
{\rm Para} &=& \{y_{ee}, y_{\mu\mu}, y_{\tau\tau}, y_{\mu e}, y_{\tau e}, \theta_R, \alpha_1, \alpha_2, \Delta m^2_{21}, \Delta m^2_{3l}, \epsilon_1, \epsilon_2 \} \,, \nonumber\\
{\rm Obs} &=& \{ \tilde{y}_e, \tilde{y}_\mu, \tilde{y}_\tau, \Delta m^2_{21}, \Delta m^2_{3l}, \theta_{12}, \theta_{23},  \theta_{13}\} \,.
\end{eqnarray}
Note that the CP-violating phase $\delta$ has been measured to be $195^{\circ}{}^{+51^\circ}_{-25^{\circ}}$ at the best-fit value $\pm 1\sigma$ errors for NO and $286^\circ{}^{+27^\circ}_{-32^\circ}$ at the best-fit value $\pm 1\sigma$ errors for IO, respectively. We have not included the it in the Obs set as its current global fit is far away from a Gaussian distribution. Instead, we treat $\delta$ as a prediction of the model. The lepton sector shares the same weightons and thus have the same parameters $\epsilon_1$ and $\epsilon_2$ in the scan. All the other free parameters are independent of those in the quark sector. In the charged lepton mass matrix, we can rotate phases of all coefficients $\{y_{ee}, y_{\mu\mu}, y_{\tau\tau}, y_{\mu e}, y_{\tau e}\}$ away without loss of generality and scan them in the range $(-1, 1)$. In the neutrino sector, as the mass matrix is explicitly the same as that genrated in \cite{King:2019vhv}, 
we follow the same parametrisation therein and left with  five free parameters, one angle $\theta_R$, two phases $\alpha_1$, $\alpha_2$, and two mass square differences $\Delta m_{21}^2$ and $\Delta m_{3l}^2$. We scan $\theta_R$ in $[0, \pi/2)$, $\alpha_{1,2}$ in $[0, 2\pi)$, and two mass square differences in their $3\sigma$ ranges. Points for $\chi^2<10$ are abandoned. Note that the two mass square differences appear in both the Para and Obs sets. Points with $\Delta m_{21}^2$ and $\Delta m_{3l}^2$ in their $3\sigma$ ranges but leading to $\chi^2>10$ are discarded as required. Prediction of mixing angles $\theta_{13}$ and $\theta_{12}$ with $\chi^2<10$ are shown in Fig.~\ref{fig:mixing}. Predictions without considering corrections from the charged lepton are listed in the figure as a comparison. A deviation of $0.02^\circ$ from the TM$_1$ mixing could be predicted. Therefore, the TM$_1$ mixing is still a very good approximation. The CP-violating phase $\delta$ is listed as a prediction of the model, as shown in Fig.~\ref{fig:delta_chisq}. The model in both mass orderings support small deviation from the maximal CP violation, $270^\circ<\delta < 305^\circ$ for $\chi^2 <10$. 
Below is a sample of parameters and predictions of observables with $\chi^2 = 4.515$ for the NO of neutrino mass ordering, 
\begin{eqnarray} \label{eq:para_lepton}
&y_{ee} = 0.366354\,,\quad
y_{\mu \mu} = 2.52094\,,\quad
y_{\tau \tau} = 0.293129\,,\quad
y_{\mu e} = -0.260999\,,\quad
y_{\tau e} = -0.828426\,, \nonumber\\
&\theta_R = 14.87^\circ\,,\quad
\alpha_1 = 119.97^\circ\,,\quad
\alpha_2 = 46.94^\circ\,, \nonumber\\
&\Delta m_{21}^2 = 7.073 \times 10^{-5} {\rm eV}^2\,,\quad
\Delta m_{3l}^2 = 2.506 \times 10^{-3} {\rm eV}^2\,,
\end{eqnarray}
and 
\begin{eqnarray}
&\tilde{y}_{e} = 1.97 \times 10^{-6}\,,\quad
\tilde{y}_{\mu} = 4.16 \times 10^{-4}\,,\quad
\tilde{y}_{\tau} = 7.07 \times 10^{-3}\,, \nonumber\\
&\theta_{12} = 34.353^\circ\,,\quad
\theta_{23} = 48.472^\circ\,,\quad
\theta_{13} = 8.522^\circ\,,
\end{eqnarray}
where $\epsilon_1$ and $\epsilon_2$ take same values as in Eq.~\eqref{eq:para_quark}. The CP-violating phase $\delta$ in this sample is predicted to be $\delta = 285.55^\circ$.

We also check the prediction of the effective neutrino mass parameter $m_{ee}$ which is under measurements of the neutrinoless double beta decay experiments. The scanned results are show in Fig.~\ref{fig:mixing} for both the NO and IO of neutrino mass ordering. In particular, the sample point with inputs in Eq.~\eqref{eq:para_lepton} predicts $m_{\rm lightest}=0.0597$~eV and $m_{ee} = 0.0537$~eV. We have numerically checked that the contribution from charged lepton mixing matrix is negligibly small. 
Experimental upper limits of $m_{ee}$, 0.061 -- 0.165, 0.078 -- 0.239, 0.075 -- 0.350, and 0.079 -- 0.180 eV, as measured by KamLAND-Zen \cite{KamLAND-Zen:2016pfg}, EXO-200 \cite{Anton:2019wmi}, CUORE \cite{Adams:2019jhp} and GERDA \cite{Agostini:2020xta} at 90\% CL, respectively,  are shown in the figure by comparison. We also include the cosmological constraints on the neutrino masses from Planck 2018 \cite{Aghanim:2018eyx}. The two vertical grey bands refer to the disfavoured region $0.12~{\rm eV} < \sum m_i < 0.60~{\rm eV}$ and very disfavoured region $\sum m_i >0.60~{\rm eV}$, respectively. Due to neutrino mass sum rule in Eq.~\eqref{eq:mass_sum_rule}, the lightest neutrino mass is constrained to be $m_{\rm lightest} \gtrsim 0.03~{\rm eV}$, leading to most of the parameter space disfavoured by the cosmological constraints. 

We emphasise that the sum rule in Eq.~\eqref{eq:mass_sum_rule} is an unavoidable consequence of 
the residual $Z_2^{SU}$ modular symmetry, resulting in nearly degenerate masses for right-handed neutrinos, and hence 
nearly degenerate masses for the light neutrinos via the seesaw mechanism. As a result, a large sum of neutrino masses is predicted by the model, along with relatively large $m_{ee}$, even for a normal ordering. This special feature makes the model eminently 
testable by neutrinoless double beta decay experiments.

\begin{figure}[t!]
\centering
\includegraphics[width=.47\textwidth]{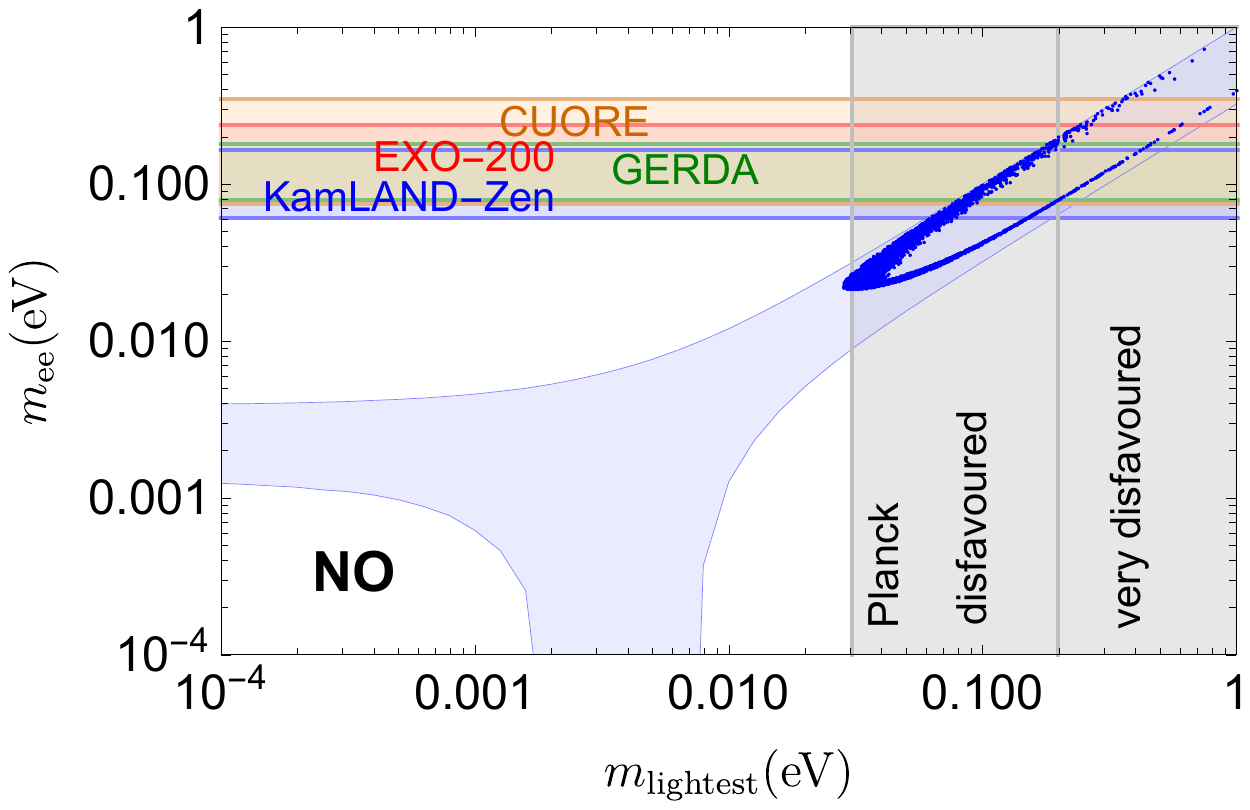}
\includegraphics[width=.47\textwidth]{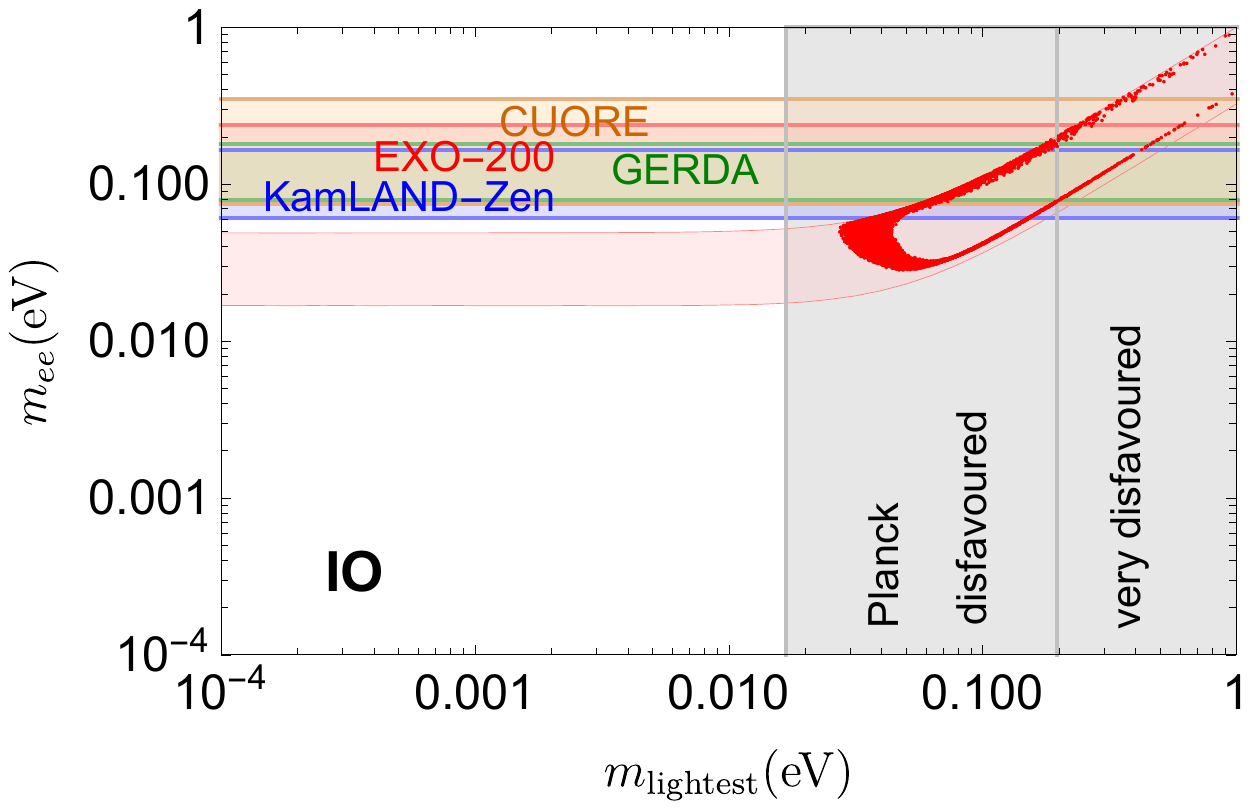}
\caption{Prediction of $m_{\rm lightest}$ vs $m_{ee}$ with $\chi^2<10$ for both normal ordering (NO, left panel) and inverted ordering (IO, right panel) of neutrino masses, allowed by the model, where $m_{\rm lightest} = m_1$ for NO and $m_{\rm lightest} = m_3$ for IO. The general parameter space of $m_{ee}$ allowed by oscillation data and current upper limit from neutrinoless double beta decay experiments KamLAND-Zen \cite{KamLAND-Zen:2016pfg}, EXO-200 \cite{Anton:2019wmi}, CUORE \cite{Adams:2019jhp} and GERDA \cite{Agostini:2020xta},  and cosmological constraints from PLANCK 2018  (disfavoured region $0.12~{\rm eV} < \sum m_i < 0.60~{\rm eV}$ and very disfavoured region $\sum m_i > 0.60~{\rm eV}$) \cite{Aghanim:2018eyx} are shown for comparison. } \label{fig:mee}
\end{figure}

\section{Conclusion}\label{sec:5}

We have constructed an $SU(5)$ GUT with twin $S_4^F \times S_4^N$ modular symmetries, accompanied by two moduli fields. This is a grand unified extension of our previous work in \cite{King:2019vhv}, allowing quark mass and mixing to be included, while preserving the 
good predictions in the lepton sector.
The two modular symmetries $S_4^F \times S_4^N$, one acting on charged fermions and one on right-handed neutrinos, are broken to a single $S_4$ symmetry by a bi-triplet scalar, leading to the effective theory invariant under one single $S_4$ but involving two modulus fields. The two modulus fields gain different VEVs, leading to the breaking of $S_4$ to a residual modular symmetry $Z_3$ in the charged fermion sectors, and a residual modular symmetry $Z_2$ in the neutrino sector, ensuring the leading order TM$_1$ lepton mixing. 

In the neutrino sector, the model has the same flavour structure as that in \cite{King:2019vhv}. In the charged fermion sector, there are two main differences from our former work: 1) charged fermion mass hierarchies are explained due to the coupling to the two weightons; 2) small mixings are induced in the charged fermion mass matrices. The triangular form of the charged lepton and down quark Yukawa matrices plays a special role in this model, ensuring suppressed charged lepton corrections to the PMNS matrix, while allowing the down quark Yukawa matrix to dominantly contribute to Cabibbo mixing, where we have shown that a good fit may be achieved to the CKM matrix. 
The model predicts TM$_1$ lepton mixing to very good approximation, and neutrinoless double beta decay at rates close to the sensitivity of current and future experiments, for both normal and inverted orderings.

\section*{Acknowledgements}
SFK and YLZ acknowledge the STFC Consolidated Grant 
ST/L000296/1 and the European Union’s Horizon 2020 Research and Innovation programme under Marie Sk\l{}odowska-Curie grant agreement HIDDeN European ITN project (H2020-MSCA-ITN-2019//860881-HIDDeN).

\appendix

\section{Modular group and modular forms of level 4}
\label{app:1}

The modular group $\overline{\Gamma}$ is a set of modular transformations acting on the upper complex plane. Given a modulus viable $\tau$ with ${\rm Im}(\tau)>0$, each element $\gamma$ of $\overline{\Gamma}$ appears as a linear fractional transformation 
\begin{eqnarray} \label{eq:modular_transformation_p}
\gamma:~ \tau \to \gamma \tau = \frac{a \tau + b}{c \tau + d}
\end{eqnarray}
with $a$, $b$, $c$, and $d$ are any integers satisfying $ad-bc=1$.
As $\gamma$ can be represented by a $2 \times 2$ matrix in Eq.~\eqref{eq:modular_transformation}, the group $\overline{\Gamma}$ is expressed as 
\begin{eqnarray}
\overline{\Gamma} = \left\{ \begin{pmatrix} a & b \\ c & d \end{pmatrix} / (\pm \mathbf{1})\,, ~~ a, b, c, d \in \mathbb{Z}, ~~  ad-bc=1  \right\} \,.
\end{eqnarray}
The modular group is an infinite group. It has two generators, $S_\tau$ and $T_\tau$, satisfying $S_\tau^2 = (S_\tau T_\tau)^3 = \mathbf{1}$. These generators act on the modulus $\tau$ as, 
\begin{eqnarray}
S_\tau:~ \tau \to -\frac{1}{\tau} \,, \quad
T_\tau:~ \tau \to \tau + 1\,,
\end{eqnarray}
respectively, which, represented as $2 \times 2$ matrices, are given by 
\begin{eqnarray} \label{eq:generators}
S_\tau=\begin{pmatrix} 0 & 1 \\ -1 & 0 \end{pmatrix}\,, \hspace{1cm}
T_\tau=\begin{pmatrix} 1 & 1 \\ 0 & 1 \end{pmatrix} \,.
\end{eqnarray}
With the requirement $a, d = 1~({\rm mod}~4)$ and $b, c =  0~({\rm mod}~4)$, a subset of $\overline{\Gamma}$, which is also infinite, is obtained, 
\begin{eqnarray}
\overline{\Gamma}(4) = \left\{ \begin{pmatrix} a & b \\ c & d \end{pmatrix} \in PSL(2,\mathbb{Z}), ~~ 
\begin{pmatrix} a & b \\ c & d \end{pmatrix} = \begin{pmatrix} 1 & 0 \\ 0 & 1 \end{pmatrix} ~~ ({\rm mod}~ 4) \right\} \,.
\end{eqnarray}
$\Gamma_4$ is the quotient group $\Gamma_4 = \overline{\Gamma}/\overline{\Gamma}(4)$. 
It is equivalently obtained by imposing the identity $T_\tau^4 = \mathbf{1}$. As a subgroup of $\overline{\Gamma}$, its elements can also be represented as $2 \times 2$ matrices, but the representation matrices are not unique. As the quotient group $\overline{\Gamma}/\overline{\Gamma}(4)$, each element $\gamma$ of $\Gamma_4$ satisfy the equality 
\begin{eqnarray} \label{eq:rep_1}
\eta \begin{pmatrix} 4 k_a +a & 4 k_b +b \\ 4 k_c +c & 4 k_d +d \end{pmatrix} = \begin{pmatrix} a & b \\ c & d \end{pmatrix}  \,,
\end{eqnarray} 
where $k_a$, $k_b$, $k_c$ and $k_d$ are integers and satisfy $4 k_a k_d + a k_d + d k_a = 4 k_b k_c + b k_c + c k_b$ and $\eta= \pm 1$. 

The finite modular group $\Gamma_4$ is isomorphic to $S_4$. 
The latter is the permutation group of 4 objects, see e.g.\ \cite{Escobar:2008vc}. It has 5 irreducible representations, $\mathbf{1}$, $\mathbf{1}'$, $\mathbf{2}$, $\mathbf{3}$ and $\mathbf{3}'$. 
In the studies of flavour symmetries, a common set of generators of $S_4$ are $S$, $T$ and $U$ which satisfy $S^2 = T^3 = U^2 = (ST)^3 = (SU)^2 = (TU)^2 = {\bf 1}$. We work in a widely used basis with representation matrices of $S$, $T$ and $U$ listed in Table~\ref{tab:rep_matrix_main}. 

\begin{table}[h!]
\begin{center}
\begin{tabular}{|c|ccc|}
\hline\hline
   & $\rho(T)$ & $\rho(S)$ & $\rho(U)$  \\\hline
$\mathbf{1}$ & 1 & 1 & 1 \\
$\mathbf{1^{\prime}}$ & 1 & 1 & $-1$ \\
$\mathbf{2}$ & 
$\left(
\begin{array}{cc}
 \omega  & 0 \\
 0 & \omega ^2 \\
\end{array}
\right)$ & 
$\left(
\begin{array}{cc}
 1 & 0 \\
 0 & 1 \\
\end{array}
\right)$ & 
$\left(
\begin{array}{cc}
 0 & 1 \\
 1 & 0 \\
\end{array}
\right)$ \\

$\mathbf{3}$ &  $\left(
\begin{array}{ccc}
 1 & 0 & 0 \\
 0 & \omega ^2 & 0 \\
 0 & 0 & \omega  \\
\end{array}
\right)$ &
$\frac{1}{3} \left(
\begin{array}{ccc}
 -1 & 2 & 2 \\
 2 & -1 & 2 \\
 2 & 2 & -1 \\
\end{array}
\right)$ &
$\left(
\begin{array}{ccc}
 1 & 0 & 0 \\
 0 & 0 & 1 \\
 0 & 1 & 0 \\
\end{array}
\right)$ \\

$\mathbf{3}'$ &  $\left(
\begin{array}{ccc}
 1 & 0 & 0 \\
 0 & \omega ^2 & 0 \\
 0 & 0 & \omega  \\
\end{array}
\right)$ &
$\frac{1}{3} \left(
\begin{array}{ccc}
 -1 & 2 & 2 \\
 2 & -1 & 2 \\
 2 & 2 & -1 \\
\end{array}
\right)$ &
$-\left(
\begin{array}{ccc}
 1 & 0 & 0 \\
 0 & 0 & 1 \\
 0 & 1 & 0 \\
\end{array}
\right)$ \\ \hline\hline

\end{tabular}
\caption{\label{tab:rep_matrix_main} Representation matrices for the $S_4$ generators $T$, $S$ and $U$ used in the paper, where $\omega=e^{2\pi i/3}$.}
\end{center}
\end{table}
%

As $S_4 \simeq \Gamma_4$, $S$, $T$ and $U$ can be represented by $S_\tau$ and $T_\tau$, and vice versa, 
\begin{eqnarray}
& T = S_\tau T_\tau \,,\qquad
S = T_\tau^2 \,,\qquad
U = T_\tau S_\tau T_\tau^2 S_\tau \,; \nonumber\\
& T_\tau = U T S T^2 \,,\qquad
S_\tau = STSU\,.
\end{eqnarray}
Any element of $S_4$ can be represented by $S_\tau$ and $T_\tau$. For example, $SU$, which is crucial to achieve the special mass structure in the neutrino sector, is given by $SU = S_\tau T_\tau S_\tau T_\tau^{-1} S_\tau$. 
Given representation matrices of $S_\tau$ and $T_\tau$ in Eq.~\eqref{eq:generators}, it is straightforward to obtain $2 \times 2$ matrices  of $S$, $T$, $U$ and $SU$ as
\begin{eqnarray} \label{eq:STU}
T=\begin{pmatrix} 0 & 1 \\ -1 & -1 \end{pmatrix} \,, \quad
S=\begin{pmatrix} 1 & 2 \\ 0 & 1 \end{pmatrix}\,, \quad
U=\begin{pmatrix} 1 & -1 \\ 2 & -1 \end{pmatrix} \,, \quad
SU=\begin{pmatrix} -1 & -1 \\ 2 & 1 \end{pmatrix}\,.
\end{eqnarray}
We emphasise that the representation matrix for each element is not unique. An alternative representation matrix is obtained by the equality in Eq.~\eqref{eq:rep_1}. 

A main difference of the modular invariance approach from the classical flavour symmetry approach is that the Yukawa couplings are formed as a consequence of modular forms, instead of combinations of a series of flavon VEVs. Modular forms are holomorphic functions of $\tau$ under modular transformations. 

Modular forms of level 4 are classified by modular weights. The latter must be even and positive integrals, which we label as $2k$. There are $4k+1$ linearly independent modular forms of level 4 and weight $2k$. They are all decomposed to irreducible representations of $S_4$. 
For $2k=2$, there are 5 modular forms. They are decomposed into a doublet $\mathbf{2}$ and a triplet $\mathbf{3}'$ of $S_4$, 
\begin{eqnarray} 
Y^{(2)}_{\mathbf{2}}(\tau) = \begin{pmatrix} Y_1(\tau) \\ Y_2(\tau) \end{pmatrix} \,,~\quad
Y^{(2)}_{\mathbf{3}'}(\tau) = \begin{pmatrix} Y_3(\tau) \\ Y_4(\tau) \\ Y_5(\tau) \end{pmatrix} \,.
\end{eqnarray}
A non-linear algebra satisfied among three of the modular forms, 
\begin{eqnarray}  \label{eq:algebra}
(Y_3^2 + 2 Y_4 Y_5)^2 = (Y_4^2 + 2 Y_3 Y_5) (Y_5^2 + 2 Y_3 Y_4)
\end{eqnarray}
is satisfied \cite{Penedo:2018nmg}. This constraint is essential to cover the modular space of $\Gamma_4$. Modular forms of higher weights are contracted from $Y_1, \cdots, Y_5$. For $2k=4$, there are 9 linearly independent modular forms, forming a singlet $\mathbf{1}$, a doublet $\mathbf{2}$ and two triplets $\mathbf{3}$ and $\mathbf{3}'$ of $S_4$, 
\begin{eqnarray} 
Y^{(4)}_{\mathbf{1}}(\tau) = Y_1Y_2 \,,&&~
Y^{(4)}_{\mathbf{2}}(\tau) = \begin{pmatrix} Y_2^2 \\ Y_1^2 \end{pmatrix}\,, \nonumber\\
Y^{(4)}_{\mathbf{3}}(\tau) = \begin{pmatrix} Y_1 Y_4 - Y_2 Y_5 \\ Y_1 Y_5 - Y_2 Y_4 \\ Y_1 Y_3 - Y_2 Y_4 \end{pmatrix}\,,&&~
Y^{(4)}_{\mathbf{3}'}(\tau) = \begin{pmatrix} Y_1 Y_4 + Y_2 Y_5 \\ Y_1 Y_5 + Y_2 Y_4 \\ Y_1 Y_3 + Y_2 Y_4 \end{pmatrix} \,.
\end{eqnarray}
For $2k=6$, the number is increased to 13. They are decomposed to
\begin{eqnarray} 
Y^{(6)}_{\mathbf{1}}(\tau) = Y_1^3+Y_2^3 \,, &&
Y^{(6)}_{\mathbf{1}'}(\tau) = Y_1^3-Y_2^3 \,, \nonumber\\
Y^{(6)}_{\mathbf{2}}(\tau) = Y^{(4)}_{\mathbf{1}} Y^{(2)}_{\mathbf{2}} \,, &&
Y^{(6)}_{\mathbf{3}',1}(\tau) = Y^{(4)}_{\mathbf{1}} Y^{(2)}_{\mathbf{3}'}\,, \nonumber\\
Y^{(6)}_{\mathbf{3}}(\tau) = \begin{pmatrix} Y_2^2 Y_4 - Y_1^2 Y_5 \\ Y_2^2 Y_5 - Y_1^2 Y_4 \\ Y_2^2 Y_3 - Y_1^2 Y_4 \end{pmatrix}\,,&&
Y^{(6)}_{\mathbf{3}',2}(\tau) = \begin{pmatrix} Y_2^2 Y_4 + Y_1^2 Y_5 \\ Y_2^2 Y_5 + Y_1^2 Y_4 \\ Y_2^2 Y_3 + Y_1^2 Y_4 \end{pmatrix} \,.
\end{eqnarray}
More modular forms with higher modular weights are listed in \cite{Penedo:2018nmg}. In particular, singlet modular forms at weights 8, 10 and 12 are respectively given by
\begin{eqnarray}
&Y^{(8)}_{\mathbf{1}}(\tau) = [Y^{(4)}_{\mathbf{1}}]^2 \,, ~
Y^{(10)}_{\mathbf{1}}(\tau) = Y^{(4)}_{\mathbf{1}}  Y^{(6)}_{\mathbf{1}} \,, ~
Y^{(10)}_{\mathbf{1}'}(\tau) = Y^{(4)}_{\mathbf{1}}  Y^{(6)}_{\mathbf{1}'} \,, ~\nonumber\\
&Y^{(12)}_{\mathbf{1},1}(\tau) = [Y^{(6)}_{\mathbf{1}}]^2 \,, ~
Y^{(12)}_{\mathbf{1},2}(\tau) =[Y^{(6)}_{\mathbf{1}'}]^2 \,, ~
Y^{(12)}_{\mathbf{1},3}(\tau) =[Y^{(4)}_{\mathbf{1}'}]^3 \,, ~
Y^{(12)}_{\mathbf{1}'}(\tau) = Y^{(6)}_{\mathbf{1}} Y^{(6)}_{\mathbf{1}'} \,.
\end{eqnarray}


A modular form at a stabiliser takes an interesting weight-dependent direction which satisfies \cite{deMedeirosVarzielas:2019cyj}
\begin{eqnarray} \label{eq:yukawa_eigenvector}
\rho_I(\gamma) Y_I(\tau_\gamma) = (c\tau_\gamma + d)^{-2k} Y_I(\tau_\gamma) \,. 
\end{eqnarray} 
Namely, a modular form at a stabiliser $\tau_\gamma$ is an eigenvector of the representation matrix $\rho_I(\gamma)$ with respective eigenvalue $(c\tau_\gamma + d)^{-2k}$. 
In the special case $(c\tau_\gamma + d)^{-2k}=1$, leading to $\rho_I(\gamma) Y_I(\tau_\gamma) = Y_I(\tau_\gamma)$, the residual modular symmetry is reduced to the residual flavour symmetry. 
Otherwise, the residual modular symmetry is different from the latter. 

We discuss which directions triplet modular forms $Y_{\mathbf{3}^{(\prime_\gamma)}}^{(2k)}(\tau)$ may take at $\tau = \tau_T$ and $\tau_{SU}$. 
The eigenvalue $(c\tau_\gamma+d)^{-2k}$ at these stabilisers is given by $(-\tau_T-1)^{-2k} = \omega^{2k}$ at $\gamma_T$ and $(2\tau_{SU}+1)^{-2k} = (-1)^{k}$ at $\gamma_{SU}$,  respectively. 
Given triplet ($\mathbf{3}$, $\mathbf{3}'$) representation matrices for $S$, $T$ and $U$ in Table~\ref{tab:rep_matrix_main}, it is straightforward to obtain 
\begin{eqnarray} \label{eq:modular_triplet}
Y^{(2)}_{\mathbf{3}^{(\prime)}}(\tau_T), Y^{(8)}_{\mathbf{3}^{(\prime)}}(\tau_T) \propto \begin{pmatrix} 0 \\ 1 \\ 0 \end{pmatrix}\,,~~
Y^{(4))}_{\mathbf{3}^{(\prime)}}(\tau_T) \propto \begin{pmatrix} 0 \\ 0 \\ 1 \end{pmatrix}\,,~~
Y^{(6))}_{\mathbf{3}^{(\prime)}}(\tau_T) \propto \begin{pmatrix} 1 \\ 0 \\ 0 \end{pmatrix}\,,~~
Y^{(2)}_{\mathbf{3}}(\tau_{SU}), Y^{(4)}_{\mathbf{3}'}(\tau_{SU})
\propto \begin{pmatrix} 2 \\ -1 \\ -1  \end{pmatrix}\,.\nonumber\\
\end{eqnarray}
These results are directly obtained from the symmetry argument, without knowing explicit expressions of modular forms. 
However, there are some exceptions of modular forms whose directions cannot be directly obtained from the above argument, e.g., $Y^{(4)}_{\mathbf{3}}(\tau_{SU})$, $Y^{(2)}_{\mathbf{3}'}(\tau_{SU})$, which correspond to eigenvectors of degenerate eigenvalues. Their directions have been calculated in \cite{King:2019vhv} with the help of the identity in Eq.~\eqref{eq:algebra}, 
\begin{eqnarray} \label{eq:modular_triplet}
Y^{(2)}_{\mathbf{3}'}(\tau_{SU}) \propto \begin{pmatrix} 1 \\ 1-\sqrt{6} \\ 1+\sqrt{6} \end{pmatrix} \,,~ 
Y^{(4)}_{\mathbf{3}}(\tau_{SU}) 
\propto \begin{pmatrix} 
 \sqrt{2} \\
 \sqrt{2}-\sqrt{3} \\
 \sqrt{2}+\sqrt{3}  \end{pmatrix}\,,
\end{eqnarray}
For the singlet modular forms, some of them may vanish at the stabilisers, e.g., $Y^{(4)}_{\mathbf{1}}(\tau_T) = Y^{(8)}_{\mathbf{1}}(\tau_T) = Y^{(10)}_{\mathbf{1}}(\tau_T) = 0$. They do not contribute to fermion masses.

\section{Vacuum alignments \label{app:2}}

We discuss how to achieve the required VEVs of the bi-triplet scalar and weightons using the general driving-field approach in the supersymmetry.

The vacuum alignment for the bi-triplet scalar $\Phi$ can be realised by introducing two driving fields
$\chi_{FN} \sim (\mathbf{3}, \mathbf{3})$ and $\chi_{N} \sim (\mathbf{1}, \mathbf{3})$
of $S_4^F \times S_4^N$ with zero modular weights. The general superpotential terms for driving fields are given by
\begin{eqnarray}
w_d &\supset& \big((\Phi \Phi)_{(\mathbf{3}, \mathbf{3})} + {\rm M}_\Phi \Phi \big) \chi_{FN} + (\Phi \Phi)_{(\mathbf{1}, \mathbf{3})}  \chi_N \,,
\end{eqnarray}
where ${\rm M}$ is a mass-dimensional coefficient. 
Minimisation of the superpotential gives rise to 
\begin{eqnarray} \label{eq:identities1}
&&(\Phi \Phi)_{(\mathbf{3}, \mathbf{3})} + {\rm M}_\Phi \Phi = 0 \,, \nonumber\\
&&(\Phi \Phi)_{(\mathbf{1}, \mathbf{3})} = 0 \,,
\end{eqnarray}
As proven in \cite{deMedeirosVarzielas:2019cyj}, these identities guarantee the VEV of $\Phi$ as in Eq.~\eqref{eq:vev} up to an unphysical basis transformation, thus leading to the breaking $S_4^F \times S_4^N \to S_4$. 

The non-vanishing VEVs of weightons can be realised by introducing two driving fields $\chi_1$ and $\chi_2$, both of which are trivial singlets of $S_4^F \times S_4^N$ with zero modular weights. Superpotential terms are
\begin{eqnarray}
w_d &\supset& \chi_1 \Big(\sum_{k=0}^4 y^{(4+2k,4)}_{1} (\tau_F,\tau_N)\frac{\phi_1^k \phi_2^{4-k}}{\Lambda^2}  - {\rm M}_1^2 \Big) 
+ \chi_2 \Big(\sum_{k=0}^4 y^{(4+2k,4)}_{2} (\tau_F,\tau_N)\frac{\phi_1^k \phi_2^{4-k}}{\Lambda^2} - {\rm M}_2^2 \Big) \,,
\end{eqnarray}
where $M_1$ and $M_2$ are two mass-dimensionful parameters, $y^{(4+2k,4)}_{1}$ and $y^{(4+2k,4)}_{1}$ are any singlet modular form of modular weights $4+2k$ and $4$ in $S_4^F \times S_4^N$ which are allowed by the symmetries. We have ignored higher dimensional terms with dimension $\geqslant 7$. The superpotential gives rise to 
\begin{eqnarray} \label{eq:identities2}
&& y^{(6,4)}_{1} (\tau_T,\tau_{SU})\frac{\phi_1 \phi_2^{3}}{\Lambda^2} + y^{(12,4)}_{1} (\tau_T,\tau_{SU})\frac{\phi_1^4}{\Lambda^2} - {\rm M}_1^2 = 0 \,, \nonumber\\
&& y^{(6,4)}_{2} (\tau_T,\tau_{SU})\frac{\phi_1 \phi_2^{3}}{\Lambda^2} + y^{(12,4)}_{2} (\tau_T,\tau_{SU})\frac{\phi_1^4}{\Lambda^2} - {\rm M}_2^2 = 0 \,,
\end{eqnarray}
where we have taken $\langle \tau_F \rangle = \tau_T$ and $\langle \tau_N \rangle = \tau_{SU}$ and used $Y^{(4)}_{\mathbf{1}}(\tau_T)=Y^{(8)}_{\mathbf{1}}(\tau_T)=Y^{(10)}_{\mathbf{1}}(\tau_T)=0$.
Eq.~\eqref{eq:identities2} assures non-trivial vacuum of $\phi_1$ and $\phi_2$. However, The correlation $\epsilon_1  \sim \sqrt{\epsilon_2} \sim \theta_C/\sqrt{2}$ cannot be derived from the identities.

\end{document}